\newcommand\apjcls{1}
\newcommand\aastexcls{2}
\newcommand\othercls{3}

\newcommand\papercls{\aastexcls}
\documentclass[tighten, times, twocolumn]{aastex62}  

\if\papercls \apjcls
\usepackage{apjfonts}
\else\if\papercls \othercls
\usepackage{epsfig}
\usepackage{margin}
\usepackage{times}
\fi\fi
\usepackage{ifthen}
\usepackage{natbib}
\usepackage{amssymb, amsmath}
\usepackage{appendix}
\usepackage{etoolbox}
\usepackage[T1]{fontenc}
\usepackage{paralist}
\if\papercls \apjcls
\newcommand\aas{\ref@jnl{AAS Meeting Abstracts}}
\newcommand\dps{\ref@jnl{AAS/DPS Meeting Abstracts}}
\newcommand\maps{\ref@jnl{MAPS}}
\else\if\papercls \othercls
\usepackage{astjnlabbrev-jh}
\fi\fi

\if\papercls \aastexcls
\hypersetup{citecolor=blue, 
            linkcolor=blue, 
            menucolor=blue, 
            urlcolor=blue}  
\else
\usepackage[
bookmarks=true,           
bookmarksnumbered=true,   
colorlinks=true,          
citecolor=blue,           
linkcolor=blue,           
menucolor=blue,           
urlcolor=blue,            
linkbordercolor={0 0 1},  
pdfborder={0 0 1},
frenchlinks=true]{hyperref}
\fi

\if\papercls \othercls
\newcommand{\eprint}[1]{\href{http://arxiv.org/abs/#1}{#1}}
\else
\renewcommand{\eprint}[1]{\href{http://arxiv.org/abs/#1}{#1}}
\fi

\providecommand{\adsurl}[1]{\href{#1}{ADS}}

\makeatletter
\patchcmd{\NAT@citex}
  {\@citea\NAT@hyper@{%
     \NAT@nmfmt{\NAT@nm}%
     \hyper@natlinkbreak{\NAT@aysep\NAT@spacechar}{\@citeb\@extra@b@citeb}%
     \NAT@date}}
  {\@citea\NAT@nmfmt{\NAT@nm}%
   \NAT@aysep\NAT@spacechar\NAT@hyper@{\NAT@date}}{}{}

\patchcmd{\NAT@citex}
  {\@citea\NAT@hyper@{%
     \NAT@nmfmt{\NAT@nm}%
     \hyper@natlinkbreak{\NAT@spacechar\NAT@@open\if*#1*\else#1\NAT@spacechar\fi}%
       {\@citeb\@extra@b@citeb}%
     \NAT@date}}
  {\@citea\NAT@nmfmt{\NAT@nm}%
   \NAT@spacechar\NAT@@open\if*#1*\else#1\NAT@spacechar\fi\NAT@hyper@{\NAT@date}}
  {}{}
\makeatother

\makeatletter
\DeclareRobustCommand{\lowcase}[1]{\@lowcase#1\@nil}
\def\@lowcase#1\@nil{\if\relax#1\relax\else\MakeLowercase{#1}\fi}
\makeatother

\DeclareSymbolFont{UPM}{U}{eur}{m}{n}
\DeclareMathSymbol{\umu}{0}{UPM}{"16}
\let\oldumu=\umu
\renewcommand\umu{\ifmmode\oldumu\else\math{\oldumu}\fi}

\if\papercls \othercls

\else

\fi

\let\oldsim=\sim
\renewcommand\sim{\ifmmode\oldsim\else\math{\oldsim}\fi}
\let\oldpm=\pm
\renewcommand\pm{\ifmmode\oldpm\else\math{\oldpm}\fi}
\newcommand\by{\ifmmode\times\else\math{\times}\fi}

\newbox{\wdbox}
\renewcommand\c{\setbox\wdbox=\hbox{,}\hspace{\wd\wdbox}}
\renewcommand\i{\setbox\wdbox=\hbox{i}\hspace{\wd\wdbox}}

\newcount\timect
\newcount\hourct
\newcount\minct
\newcommand\now{\timect=\time \divide\timect by 60
         \hourct=\timect \multiply\hourct by 60
         \minct=\time \advance\minct by -\hourct
         \number\timect:\ifnum \minct < 10 0\fi\number\minct}

\catcode`@=11

\newcommand\comment[1]{}

\newcommand\commenton{\catcode`\%=14}

\renewcommand\math[1]{$#1$}
\newcommand\mathshifton{\catcode`\$=3}

\let\atab=&
\newcommand\atabon{\catcode`\&=4}

\let\oldmsp=\sp
\let\oldmsb=\sb
\def\sp#1{\ifmmode
           \oldmsp{#1}%
         \else\strut\raise.85ex\hbox{\scriptsize #1}\fi}
\def\sb#1{\ifmmode
           \oldmsb{#1}%
         \else\strut\raise-.54ex\hbox{\scriptsize #1}\fi}
\newbox\@sp
\newbox\@sb
\def\sbp#1#2{\ifmmode%
           \oldmsb{#1}\oldmsp{#2}%
         \else
           \setbox\@sb=\hbox{\sb{#1}}%
           \setbox\@sp=\hbox{\sp{#2}}%
           \rlap{\copy\@sb}\copy\@sp
           \ifdim \wd\@sb >\wd\@sp
             \hskip -\wd\@sp \hskip \wd\@sb
           \fi
        \fi}
\def\msp#1{\ifmmode
           \oldmsp{#1}
         \else \math{\oldmsp{#1}}\fi}
\def\msb#1{\ifmmode
           \oldmsb{#1}
         \else \math{\oldmsb{#1}}\fi}

\def\supon{\catcode`\^=7}

\def\subon{\catcode`\_=8}

\def\supsubon{\supon \subon}

\newcommand\actcharon{\catcode`\~=13}

\newcommand\paramon{\catcode`\#=6}

\comment{And now to turn us totally on and off...}

\newcommand\reservedcharson{ \commenton  \mathshifton  \atabon  \supsubon 
                             \actcharon  \paramon}

\catcode`@=12
\reservedcharson

\if\papercls \apjcls

\else

\fi

\newcommand\chisq{\ifmmode{\chi\sp{2}}\else\math{\chi\sp{2}}\fi}
\newcommand\redchisq{\ifmmode{ \chi\sp{2}\sb{\rm red}}
                    \else\math{\chi\sp{2}\sb{\rm red}}\fi}
\newcommand\Teq{\ifmmode{T\sb{\rm eq}}\else$T$\sb{eq}\fi}
\newcommand\mjup{\ifmmode{M\sb{\rm Jup}}\else$M$\sb{Jup}\fi}
\newcommand\rjup{\ifmmode{R\sb{\rm Jup}}\else$R$\sb{Jup}\fi}
\newcommand\msun{\ifmmode{M\sb{\odot}}\else$M\sb{\odot}$\fi}
\newcommand\rsun{\ifmmode{R\sb{\odot}}\else$R\sb{\odot}$\fi}
\newcommand\mearth{\ifmmode{M\sb{\oplus}}\else$M\sb{\oplus}$\fi}
\newcommand\rearth{\ifmmode{R\sb{\oplus}}\else$R\sb{\oplus}$\fi}

\shorttitle{Conditional‌ moments of  the first derivative in the redshift space}
\shortauthors{Jalali Kanafi \&  Movahed}
\begin{document}

 \title{Probing the Anisotropy and Non-Gaussianity in the Redshift Space through the	Conditional Moments of the First Derivative}

\author{M. H. Jalali Kanafi}
\email{m\_jalalikanafi@sbu.ac.ir}
\affiliation{ Department of Physics, Shahid Beheshti University, 1983969411, Tehran, Iran}

\author{S. M. S. Movahed}
\email{m.s.movahed@ipm.ir}
\affiliation{ Department of Physics, Shahid Beheshti University, 1983969411, Tehran, Iran}
\affiliation{School of Astronomy, Institute for Research in Fundamental Sciences (IPM), P. O. Box 19395-5531, Tehran, Iran}




\begin{abstract}
        
        Focusing on the redshift space observations with plane-parallel approximation and relying on the rotational dependency of the general definition of excursion sets, we introduce the so-called conditional moments of the first derivative  ($cmd$) measures for the smoothed matter density field in three dimensions. We derive the perturbative expansion of $cmd$ for the real space and redshift space where peculiar velocity disturbs the galaxies’ observed locations. Our criteria can successfully recognize the contribution of linear Kaiser and Finger-of-God effects. Our results demonstrate that the $cmd$ measure has significant sensitivity for pristine constraining the redshift space distortion parameter $\beta=f/b$ and interestingly, the associated normalized quantity in the Gaussian linear Kaiser limit has only  $\beta$ dependency.Implementation of the synthetic anisotropic Gaussian field approves the consistency
        between the theoretical and numerical results. Including the first-order contribution of non-Gaussianity perturbatively in the  $cmd$criterion implies that the N-body simulations for the Quijote suite in the redshift space have been mildly skewed with a higher value for the threshold greater than zero. The non-Gaussianity for the perpendicular direction to the line of sight in the redshift space for smoothing scales $R\gtrsim 20$ Mpc h$^{-1}$ is almost the same as the real space. In contrast, the non-Gaussianity along the line of sight direction in redshift space is magnified. The Fisher forecasts indicate an almost significant enhancement in constraining the cosmological parameters, $\Omega_m$, $\sigma_8$, and $n_s$ when using $cmd+cr$ jointly. 
\end{abstract}


\section{Introduction}
\label{introduction}

In the high-precision cosmology era, drastic attention should be paid to the various robust measures construction for extracting information from random cosmological fields as accurate as possible, particularly from large-scale structures of the matter distribution in the Universe \citep{peebles2020large,Bernardeau:2001qr}. On the other hand, discrepancies between what we observe through various surveys and theoretical counterparts essentially persuade researchers  to include the stochastic notion \citep{kaiser1984spatial,Bardeen:1985tr,Bernardeau:2001qr,matsubara2003statistics,codis2013non,matsubara2020statistics}.  It is supposed that on the sufficiently large scales, the distribution of galaxies in the real space is homogeneous and isotropic, while, such an assumption is no longer satisfied in the redshift space when the position of structures is plotted as a function of redshift rather than their distances. Dealing with imposed anisotropy requires designing proper methods which are sensitive to both the existence of preferred direction and non-Gaussianity generated form different mechanisms.     

The observed redshifts of galaxies which are mainly originated by the Hubble flow are also disturbed by their peculiar velocity along the line of sight. In the vicinity of peculiar velocity which is almost produced by inhomogeneity known as overdensities and underdensities in the local Universe, a difference between galaxies's actual locations and their observed locations as determined by their redshifts exists. This phenomenon is known  as the redshift-space distortion (RSD). The  Finger-of-God  (FoG) effect \citep{Jackson1972,peebles2020large} and the linear Kaiser effect \citep{Kaiser1987} are the different parts of RSD dominated in the small enough and large scales, respectively. The elongation of clusters along the line of sight caused by the random motion of galaxies within the virialized clusters on small scales is so-called FoG, while the linear Kaiser effect refers to the suppression in the clustering of galaxies on large scales due to the coherent motion into the overdense regions of density field leading to squash the shape of clusters in redshift space along the line of sight direction \citep{sargent1977statistical,Hamilton1992,hamilton1998linear}. Although RSD makes the interpretation of observational data more challenging, it provides an opportunity to extract statistical information to constrain associated cosmological parameters \citep{Hamilton1992,hamilton1998linear,Bernardeau:2001qr,Weinberg2013}.

In recent years, many researches have been focused on the analysis of RSD from different points of view. As illustration: the correlation between the redshift distortions and cosmic mass distribution makes sense to utilize the RSD for assessing the linear growth rate of density fluctuations \citep{hamaus2022euclid,panotopoulos2021growth},  trying to break the degeneracy between various modified gravity and General Relativity in the presence of massive neutrinos in the context of standard model of cosmology \citep{Wright2019}; the joint analysis of the Alcock-Paczynski effect and RSD to probe the cosmic expansion \citep{Song2015}; combining RSD  with weak lensing and baryon acoustic oscillations to improve the observational constraints on the cosmological parameters \citep{eriksen2015combining}; quantifying the RSD spectrum \citep{bharadwaj2020quantifying,mazumdar2020quantifying,mazumdar2023quantifying}; examining the primordial non-Gaussianity via RSD \citep{tellarini2016galaxy}; disentangling redshift-space distortions and non-linear bias \citep{2016MNRAS.457.1076J}.

As of the importance role of the large scale structures and corresponding observational catalogs, there are many attempts incorporating the geometrical and topological virtues of diverse relevant fields such as the genus statistic \citep{gott1986sponge,hamilton1986topology}, contour statistics including 1-, 2- and 3-Dimensional features  \citep{ryden1988area,Ryden:1988rk}, Minkowski functionals consisting of  $d+1$ scalar quantities which describe the morphology of isodensity contours of $d$-dimensional field \citep{mecke1993robust,schmalzing1997beyond} (see also  \cite[and references therein]{kerscher1997minkowski,sahni1998shapefinders,hikage2006primordial,einasto2011sloan,liu2020neutrino,liu2022probing, matsubara2003statistics,pogosyan2009local
	,gay2012non,codis2013non,matsubara2021weakly,matsubara2022minkowski}).

The central assumptions in many  cosmological studies are homogeneity, isotropy, and Gaussianity due to the extension of the central limit theorem domain (see also the \cite{2023CQGra..40i4001K} for a comprehensive explanation of Cosmological Principle).  In the real data sets, not only the violation of Gaussianity is expected, but also the anisotropy can emerge due to different reasons ranging from initial conditions, and phase transitions to the non-linearity among the evolution \citep{ade2014plancknon,2016A&A...594A..17P,Renaux-Petel:2015bja,Ade:2015hxq,2009MNRAS.396.1273H,Springel06,Bernardeau:2001qr,Ade:2013xla,2021MNRAS.503..815V}.     
Subsequently, to explore the large scale structures in the redshift space as the counterpart of the real space, many powerful statistical measures have been considered by concentrating on the non-Gaussianity and anisotropy \citep{matsubara1996statistics,codis2013non,appleby2018minkowski,appleby2019ensemble,Appleby2023}.  Recently, Minkowski tensors, an extension of scalar Minkowski functionals  \citep{ McMullen1997,Alesker1999,Beisbart2002,Hug2007TheSO,santalo2004integral, 2010JSMTE..11..010K,2013NJPh...15h3028S}, have been employed on 2- and 3-Dimensional cosmological fields in the real and redshift spaces \citep{2017JCAP...06..023G,chingangbam2017tensor,appleby2018minkowski, appleby2019ensemble,Appleby2023,goyal2021local,ApplebySDSS}.

Motivated to examine the anisotropy, asymmetry and non-Gaussianity induced in many cosmological random fields, simultaneously, we pursue the mainstream of theoretical measures construction to explore the anisotropy and non-Gaussianity and to quantify the statistical features of a generic field such as density field with $z-$anisotropic behavior in plane-parallel approximation. When the anisotropy and non-Gaussianity are interested, we advocate the utilizing of measures specially designed  to declare the anisotropy rather than using those measures such as Minkowski Functional and contouring analysis which are not in principle directional tools, however, they can recognize the anisotropy and non-Gaussianity, because they generally have the imprint of directional averaging and they may give the spurious results. 

It is worth noting that, to introduce a feasible measure, we should notice the following general properties which are necessary to achieve proper cosmological inferences: it should be robust as much as possible against numerical uncertainties and finite size sampling effect. Since we are interested in using the new measure to constrain the cosmological parameters, another aspect that should be taken into account is that it is possible to establish analytical or semi-analytical prediction for the introduced measure, however, in the absence of theoretical prediction for the desired measure, there are some approaches to overwhelm this issue such as Gaussian Processes Regression \citep{2020arXiv200910862W}.

The novelties and advantages of our approach are as follows: \\
(1) We will provide a comprehensive mathematical description of the so-called {\it conditional moments of the first derivative} ($cmd$) of the fields corresponding to the excursion set and calculate the theoretical prediction of this statistic for a $3$-Dimensional isotropic and asymmetric Gaussian field as a function of threshold using a probabilistic framework. We will also take into account the first order correction due to the mildly non-Gaussianity in the context of perturbative approach. Our notable measure is able to recognize the preferred and generally anisotropic directions for any generic field for 2- and 3-Dimension in different disciplines as well as non-Gaussianity \citep{li2013detection,nezhadhaghighi2015crossing,klatt2022characterization,2010JSMTE..11..010K,2013NJPh...15h3028S}. \\
(2) The anisotropy imprint by the linear Kaiser effect will be examined by our introduced measure as well as crossing statistics as a particular generalization of Minkowski Functionals in the plane-parallel approximation. Also incorporating the Gaussian and Lorentzian phenomenological models of the FoG effect, the correction to the linear Kaiser limit will be carried out. To make our analysis more complete, we will compare the sensitivity of this statistic to the redshift space distortions parameter concerning other famous measures such as crossing statistics $(cr)$ and Minkowski tensors.\\
3) Using the N-body simulations provided by the Quijote suite, the capability of $cmd$ and $cr$ statistics will be verified and we will elucidate the non-Gaussianity matter density field in redshift space, especially thorough the line of sight by $cmd$ up to the $\mathcal{O}(\sigma^2_0)$, perturbatively.\\
4) By performing Fisher forecasts, we evaluate the power of $cmd$ and $cr$ to constrain the relevant cosmological parameters. \\
5) The sensitivity of $cmd$ and $cr$ to the halo bias will be examined by Quijote halo catalogs in redshift space.      

The rest of this paper is organized as follows: Section \ref{sec:RSD} will be assigned to a brief review the notion of RSD and the relationship between the density field in the redshift and real spaces. In Section \ref{section:Probabilistic Framework}, we will present a mathematical description of our new measure to capture the preferred direction in the context of a probabilistic framework. The perturbative expansion of theoretical prediction for the $cmd$ in the mildly non-Gaussian regime is also given in this section.  Section \ref{sec:IRSD}  will be devoted to the characterization of RSD including the linear Kaiser and FoG effects using the geometrical measures. The implementation of geometrical measures, $cmd$ and $cr$ on our synthetic data sets and also N-body simulations by the Quijote team will be presented in section \ref{sec:mockdata}. We will give the results of Fisher forecasts and also halo bias dependency and sensitivity, in this section. The last section will be focused on the summary and concluding remarks.

\section{Redshift Space Distortions}\label{sec:RSD}
In this section, for the sake of clarity, we first briefly review the relationship between a typical cosmological stochastic field in the redshifted Universe and corresponding quantity in the real space. Owing to the peculiar velocity field, the observed position of an object in redshift space ($\boldsymbol{s}$) differs from its real space position, $\boldsymbol{r}$, and its relation is given by:
\begin{eqnarray}
	\boldsymbol{s} = \boldsymbol{r} + \frac{\boldsymbol{v(\boldsymbol{r})}.\hat{\boldsymbol{n}} }{H} \hat{\boldsymbol{n}},
	\label{eq:redshift position}
\end{eqnarray}
where $\boldsymbol{v(\boldsymbol{r})}$ represents peculiar velocity, $\hat{\boldsymbol{n}}$ is the line of sight direction and $H$ is the Hubble parameter. Equation   (\ref{eq:redshift position}) leads to a distortions in an observed cosmological stochastic field, particularly the observed density field in redshift space. To the linear order, due to the so-called linear Kaiser effect, the distorted density contrast field in redshift space is related to the density contrast field in real space for a given wavenumber, $\boldsymbol{k}$, by \citep{Kaiser1987}:
\begin{eqnarray}
	\tilde{\delta}^{(s)}(\boldsymbol{k}) = (1+\beta\mu^2)\;\tilde{\delta}^{(r)}(\boldsymbol{k})\;, \hspace{0.5cm} \beta \equiv f/b
	\label{eq:linear kaiser}
\end{eqnarray}
The (\textasciitilde) symbol is reserved for quantity in the Fourier space throughout this paper. The $\diamond$ is replaced by ($s$) and ($r$) for redshift and real spaces, respectively. Also $\mu\equiv\hat{\boldsymbol{k}}.\hat{\boldsymbol{n}}$, $f$ is the linear
growth rate of the density contrast and $b$ is the linear bias factor. Equation   (\ref{eq:linear kaiser}) holds for the matter and biased tracers (e.g. galaxies) density fields, which for the matter case, we have $b = 1$.

Beside to the linear Kaiser effect, there are non-linear effects such as the non-linear Kaiser effect and the FoG effect leading to the distortions of the density field in redshift space with different manners. Therefore, taking into account the nonlinear effects, the Equation   (\ref{eq:linear kaiser}) can be written in the general form as:
\begin{eqnarray}
	\tilde{\delta}^{(s)}(\boldsymbol{k}) = \tilde{O}_{s}(\mu,k\mu)\tilde{\delta}^{(r)}(\boldsymbol{k}).
	\label{eq:anisotropic delta}
\end{eqnarray}
in which the operator $\tilde{O}_{s}$ can be written in the multiplication decomposition of the linear Kaiser part ($\tilde{O}_{\rm{lin}}$) and the non-linear part ($\tilde{O}_{\rm{nl}}$) as below:
\begin{eqnarray}
	\tilde{O}_{s}(\mu,k\mu)& =& \tilde{O}_{\rm {lin}}(\mu,k\mu)\times\tilde{O}_{\rm {nl}}(\mu,k\mu)\nonumber\\
	&=&(1+\beta\mu^2)\tilde{O}_{\rm {nl}}(\mu,k\mu).
	\label{eq:operator}
\end{eqnarray}
Accordingly, the power spectrum in the redshift space and in the real space have the following relation:
\begin{eqnarray}
	P^{(s)}(\boldsymbol{k}) = (1+\beta\mu^2)^2\big|\tilde{O}_{\rm{nl}}(\mu,k\mu)\big|^2 P^{(r)}(\boldsymbol{k}),
	\label{eq:power spectrum}
\end{eqnarray}
Equations   (\ref{eq:anisotropic delta}) and   (\ref{eq:power spectrum}) demonstrate that the Fourier transform of the redshift space density field as well as the power spectrum depends on the direction of wavenumber relative to the line of sight. In other words, the density field in the redshift space is anisotropic and there is an alignment in the line of sight. In this case, we expect that a proper directional statistical measure is capable to distinguish the line of sight direction from the perpendicular directions. It turns out that the mentioned difference should be depended on the amount of anisotropy which is produced by redshift space distortions and even on the sensitivity of the considered directional statistics. For an isotropic density field, there is no difference between various directions.
As mentioned in the introduction, any conceivable measure to extract reliable cosmological results should be taken into account such  generated anisotropy which is inevitable for the astrophysical context. For this purpose, we will rely on the probabilistic framework in the next section to construct new directional statistical measures and evaluate the corresponding capabilities for desired applications.

\section{Probabilistic Farmework}\label{section:Probabilistic Framework}
Suppose that $\delta_{R}^{(r,s)}$ denotes the density field contrast in the real and redshift spaces and it is already smoothed by a smoothing window function, $W_R$, in the Fourier space as:
\begin{eqnarray}
	\tilde{\delta}^{(r,s)}_R (\boldsymbol{k})= \tilde{W}(kR)\tilde{\delta}^{(r,s)}(\boldsymbol{k}),
	\label{eq:field}
\end{eqnarray}
We define a so-called set for mentioned smoothed density field in 3-Dimension including the field itself and corresponding first derivative as  $\mathcal{A}^{(r,s)}\equiv \left\{\delta^{(r,s)},\delta_{,x}^{(r,s)},\delta_{,y}^{(r,s)},\delta_{,z}^{(r,s)}\right\}$ and for simplicity, we have omitted the subscript smoothing scale denoted by $R$ and hereafter, the superscript $(r,s)$ of $\mathcal{A}$ is dropped. In addition $\delta_{,i}^{(r,s)}\equiv \boldsymbol{\nabla}_i\delta^{(r,s)}$ and $i$ gets the $x,y,z$, representing the axises in the Cartesian coordinate.
\subsection{JPDF of Random Field}\label{section:Probabilistic Framework1}
The general form of joint probability density function (JPDF) of the set $\mathcal{A}$ including $4$ elements for the redshift space and real space, separately can be expressed by \citep{matsubara2003statistics}:
\begin{eqnarray}
	\begin{split}
		\mathcal{P}(\mathcal{A}) & =  \exp\bigg\lbrack \sum_{j=3}^{\infty}\frac{(-1)^j}{j!}\bigg (
		\sum_{\mu_1=1}^{N=4}...\sum_{\mu_j=1}^{N=4}\mathcal{K}^{(j)}_{\mu_1,\mu_2,...,\mu_j}
		\\
		& \times \frac{\partial^j}{\partial \mathcal{A}_{\mu_1} ... \partial \mathcal{A}_{\mu_j}} \bigg )\bigg\rbrack \mathcal{P}_G(\mathcal{A})
	\end{split}
\end{eqnarray}\label{eq1}
where $\mathcal{K}^{(n)}_{\mu_1,\mu_2,...,\mu_n}\equiv\left\langle \mathcal{A}_{\mu_1}\mathcal{A}_{\mu_2}...\mathcal{A}_{\mu_n} \right\rangle_c$ represents cumulant and $\mathcal{P}_G(\mathcal{A})$ is the multivariate Gaussian JPDF of the $\mathcal{A}$ and it is given by:
\begin{eqnarray}
	\mathcal{P}_G(\mathcal{A}) = \frac{\exp \left(-\frac12
		\mathcal{A}^T{\cdot} \left[\mathcal{K}^{(2)}\right]^{-1}{\cdot}\mathcal{A} \right)}{2\pi\; \sqrt{\det \mathcal{K}^{(2)}}} 
\end{eqnarray}
where $\mathcal{K}^{(2)}\equiv \left\langle \mathcal{A} \otimes \mathcal{A} \right\rangle_c$ is the $4\times4$ covariance
matrix of  $\mathcal{A}$ known as second cumulant and $\langle\rangle_c$ denotes to connected moment.  The matrix form of  $\mathcal{K}^{(2)}$ can be expressed as:
\begin{widetext}
	\begin{eqnarray}
		\begin{split}
			&\mathcal{K}^{(2)} =&  &\left(\begin{matrix}
				\vspace{0.2cm}
				\left\langle\delta^{(r,s)}\delta^{(r,s)}\right\rangle_c & \left\langle\delta^{(r,s)}\delta_{,x}^{(r,s)}\right\rangle_c & \left\langle\delta^{(r,s)}\delta_{,y}^{(r,s)}\right\rangle_c & \left\langle\delta^{(r,s)}\delta_{,z}^{(r,s)}\right\rangle_c \\
				\vspace{0.2cm}
				\left\langle\delta_{,x}^{(r,s)}\delta^{(r,s)}\right\rangle_c & \left\langle\delta_{,x}^{(r,s)}\delta_{,x}^{(r,s)}\right\rangle_c & \left\langle\delta_{,x}^{(r,s)}\delta_{,y}^{(r,s)}\right\rangle_c & \left\langle\delta_{,x}^{(r,s)}\delta_{,z}^{(r,s)}\right\rangle_c  \\
				\vspace{0.2cm}
				\left\langle\delta_{,y}^{(r,s)}\delta^{(r,s)}\right\rangle_c & \left\langle\delta_{,y}^{(r,s)}\delta_{,x}^{(r,s)}\right\rangle_c & \left\langle\delta_{,y}^{(r,s)}\delta_{,y}^{(r,s)}\right\rangle_c & \left\langle\delta_{,y}^{(r,s)}\delta_{,z}^{(r,s)}\right\rangle_c  \\
				\vspace{0.2cm}
				\left\langle\delta_{,z}^{(r,s)}\delta^{(r,s)}\right\rangle_c & \left\langle\delta_{,z}^{(r,s)}\delta_{,x}^{(r,s)}\right\rangle_c & \left\langle\delta_{,z}^{(r,s)}\delta_{,y}^{(r,s)}\right\rangle_c & \left\langle\delta_{,z}^{(r,s)}\delta_{,z}^{(r,s)}\right\rangle_c
			\end{matrix}\right)&
		\end{split}
	\end{eqnarray}
\end{widetext}
Using the notation $ i,j \in \{x,y,z\} $, the various components in the $\mathcal{K}^{(2)}$ becomes:
\begin{eqnarray}
	\left\langle\delta^{(r,s)}\delta^{(r,s)}\right\rangle_c &=& \left(\sigma_{0}^{(r,s)}\right)^2\nonumber\\    \left\langle\delta^{(r,s)}\delta_{,i}^{(r,s)}\right\rangle_c &=& 0\nonumber\\
	\left\langle\delta_{,i}^{(r,s)}\delta_{,j}^{(r,s)}\right\rangle_c &=&  \left(\sigma_{1i}^{(r,s)}\right)^2\delta_{ij}
	\label{eq:matrix components}
\end{eqnarray}
where $\delta_{ij}$ is Kronecker delta function. In the above Equation, $\sigma_m^2$ illustrates the $m$-order of spectral index and according to the power spectrum of the density field smoothed on the scale $R$ with a given window function, it reads as:
\begin{eqnarray}\label{eq:spectral0}
	&&\left(\sigma^{(r,s)}_{m}(R)\right)^2 = \int \frac{d^3\boldsymbol{k}}{(2\pi)^3} |\boldsymbol{k}|^{2m} P^{(r,s)}(\boldsymbol{k})\;\tilde{W}^2(kR)
	\end{eqnarray}
and the spectral index for derivative is:
\begin{eqnarray}\label{eq:spectral00}
&&\left(\sigma^{(r,s)}_{1i}(R)\right)^2 \equiv \bigg \langle \left(\delta^{(r,s)}_{,i} \right)^2 \bigg \rangle_c\nonumber\\
&&\quad\quad\;\;\;\;\;\;\quad\quad = \int \frac{d^3\boldsymbol{k}}{(2\pi)^3} k_i^{2} P^{(r,s)}(\boldsymbol{k})\;\tilde{W}^2(kR)
\end{eqnarray}
Accordingly, we have:
\begin{eqnarray}\label{eq:spectral1}
	\begin{aligned}
		\left(\sigma_{1}^{(r,s)}\right)^2 & = \bigg\langle\left(\delta^{(r,s)}_{,x}\right)^2+\left(\delta^{(r,s)}_{,y}\right)^2+\left(\delta^{(r,s)}_{,z}\right)^2\bigg\rangle_c \\ &=\left(\sigma_{1x}^{(r,s)}\right)^2 + \left(\sigma_{1y}^{(r,s)}\right)^2 + \left(\sigma_{1z}^{(r,s)}\right)^2,
	\end{aligned}
\end{eqnarray}
and for the isotropic 3-Dimensional field in the real space, we obtain:
\begin{eqnarray}
	\sigma_{1x}^{(r)} = \sigma_{1y}^{(r)} = \sigma_{1z}^{(r)} = \frac{\sigma_{1}^{(r)}}{\sqrt{\;3}}
\end{eqnarray} 
The observable quantity of any statistical measure, $\mathcal{F}(\mathcal{A})$, depending on the  $\mathcal{A}$, can be expressed by the following expectation value:
\begin{eqnarray}
	\begin{split}
		\langle \mathcal{F}(\mathcal{A})\rangle  &= \int d\mathcal{A}\; \mathcal{P}(\mathcal{A})\;\mathcal{F}(\mathcal{A}) \\
		&= \bigg\langle\exp\bigg\lbrack \sum_{j=3}^{\infty}\frac{1}{j!} \bigg (
		\sum_{\mu_1}^{4}...\sum_{\mu_j}^{4}\mathcal{K}^{(j)}_{\mu_1,\mu_2,...,\mu_j}
		\\
		&\times \frac{\partial^j}{\partial \mathcal{A}_{\mu_1} ... \partial \mathcal{\mathcal{A}}_{\mu_j}} \bigg ) \bigg \rbrack \mathcal{F}(\mathcal{A}) \bigg\rangle_G
	\end{split}\label{eq:expected value}
\end{eqnarray}
where
$\big\langle X\big\rangle_G \equiv \int d\mathcal{A}\; \mathcal{P}_G(\mathcal{A})X$. Therefore, in the presence of non-Gaussianity, one can obtain the statistical expectation value of $\mathcal{F}(A)$ in terms of Gaussian integrations based on perturbative formalism.

\subsection{The $cmd$ statistical measures}
For a 3-Dimensional density field with total volume $V$‌ sampled on lattice $\mathcal{M}$, we define excursion set $Q_{\vartheta}$ as a set of all field points which satisfy condition $ \delta^{(r,s)}(\boldsymbol{r}) \ge \vartheta\sigma_0^{(r,s)}$,‌‌
namely, $\mathcal{Q}_{\vartheta} = \left\{ \boldsymbol{r} \in \mathcal{M}\; |\;  \delta^{(r,s)}(\boldsymbol{r}) \ge \vartheta\sigma_0^{(r,s)}\right\}$. The boundary of mentioned excursion set, denoted by $\partial Q_{\vartheta}$, characterizes the isodensity contours of the density field at threshold $\vartheta$. 

As mentioned in the introduction, scalar MFs have been used to characterize the morphology of density contrast field. Focusing on the anisotropy imposed by various phenomena, proper measures which are designed for anisotropy detection are recommended to use. As an illustration, the redshift space distortion affects the isodensity contours of cosmological density fields with different manner in the along and perpendicular to the line of sights, consequently modification of MFs such as so-called Minkowski Valuations (MVs) (see Appendix A for more details) can be proper measures to examine such effect.  The rank-2 MVs  have been used to asses the anisotropy properties of redshift space and also distortions parameter \citep{matsubara1996genus,codis2013non,appleby2018minkowski,appleby2019ensemble,Appleby2023}.  
 The Genus and contour crossing in various dimensions has been examined in redshift space \citep{matsubara1996statistics,codis2013non}. Interestingly, those statistics revealing the one- and two-Dimensional slices depend on anisotropy due to peculiar velocities in redshift space. Consequently, we are  persuaded that other criteria similar to the well-known measures introduced for
the characterization of morphology may have the potential for anisotropy evaluation in the cosmological stochastic field.
After introducing the so-called level crossing as a powerful tool for quantifying a typical stochastic time series by S. O. Rice \citep{rice44a,rice44b}, the generalized form of that means including the Up-, down- and the conditional crossing statistics have been utilized as complementary methods for diverse applications \citep{Bardeen:1985tr,Bond:1987ub,ryden1988area,Ryden:1988rk,matsubara1996statistics,brill2000brief,matsubara2003statistics,movahed2011level,nezhadhaghighi2015crossing,klatt2022characterization}. Particularly, the contour crossing statistic corresponds to the mean number of intersections between the isodensity contours of the density field at threshold $\vartheta$ , $\partial Q_v$, and a straight line in a specific direction \citep{ryden1988area}. 
The crossing statistics is given by a specific choose  $\mathcal{G}=\delta_D\left(\delta^{(r,s)}-\vartheta\sigma^{(r,s)}_{0}\right)\;\bigg| \delta_{,i}^{(r,s)}\bigg|$, (see Equation (\ref{eq:MVs}) in Appendix A) leading to:
\begin{eqnarray}\label{eq:Ncr5}
N_{cr}^{(r,s)}(\vartheta,i;n) &=& \frac{1}{V} \int_{V} dV\; \delta_D\left(\delta^{(r,s)}-\vartheta\sigma^{(r,s)}_{0}\right)\; \left|\delta_{,i}^{(r,s)}\right|\nonumber\\ 
& = &\frac{1}{V} \int_{\partial Q_v}dA\; \frac{\left|\delta_{,i}^{(r,s)}\right|}{\left|\boldsymbol{\nabla}\delta^{(r,s)}\right|}\; 
\end{eqnarray}
Using Equations (\ref{eq:expected value}) and (\ref{eq:Ncr5}), the crossing  statistic ($cr$) for a Gaussian 3-Dimensional field can be expressed  as \citep{ryden1988area,matsubara1996statistics,matsubara2003statistics,codis2013non}:
\begin{eqnarray}\label{eq:Ncr1}
	\left\langle N_{cr}^{(r,s)}(\vartheta,i)\right\rangle_G  &=&\frac{\sigma_{1i}^{(r,s)}}{\pi\sigma_{0}^{(r,s)}}\;e^{-\vartheta^2/2} , 
\end{eqnarray}
where $i$ represents the direction of a straight line. 

To establish a new tool to quantify the directional dependency of a typical anisotropic field in the context of generalization of the MFs, some options exist incorporating relaxing the ``{\it Hadwiger's theorem}'', inspired by the crossing statistic a straightforward selection which is proper for cosmological interpretations is, $\mathcal{G}=\delta_D\left(\delta^{(r,s)}-\vartheta\sigma^{(r,s)}_{0}\right)\left(\delta_{,i}^{(r,s)}\right)^n$ called the {\it conditional moments of the first derivative ($cmd$)}. According to the definition of  characteristic function as $\mathcal{Z}_{\mathcal{A}}(\Lambda)=\int d\mathcal{A}\mathcal{P}(\mathcal{A})\rm{exp}({\mathbf{i}\Lambda.\mathcal{A}})$, we can also generate various orders of cumulants in addition to the moments and the same analysis can be done in a straightforward manner. This modification in the weight of the first partial derivative enables us to capture the footprint of anisotropic e.g. due to RSD. Selecting the regions satisfying the condition given by $\delta_D\left(\delta^{(r,s)}-\vartheta\sigma^{(r,s)}_{0}\right)$ ‌‌ and by fixing a direction, $i$, the $n$th moment of the first derivative of the fields in such direction for the captured regions to be computed.  ‌
From the theoretical aspect to define a new criterion, as we will show further, an analytical form exits for the $cmd$ measure to make a well-defined relation to desired cosmological parameters. The mathematical description of the $cmd$ criterion for $3-$Dimensional density field can be clarified as follows:
\begin{eqnarray}\label{eq:Ncmd}
\begin{split}
	N_{cmd}^{(r,s)}(\vartheta,i;n) &\equiv\frac{1}{V} \int_{\partial Q_v} dA\; \frac{\left(\delta_{,i}^{(r,s)}\right)^n}{\left|\boldsymbol{\nabla}\delta^{(r,s)}\right|}\\
	& = 	\frac{1}{V} \int_{V}dV\;  \delta_D\left(\delta^{(r,s)}-\vartheta\sigma^{(r,s)}_{0}\right)\; \left(\delta_{,i}^{(r,s)}\right)^n\; 
\end{split}
\end{eqnarray}
\newline
we utilize the surface to ‌‌volume integral transformation
\citep{schmalzing1997beyond}. About selecting a typical integrand among various options as mentioned in the appendix, we must point out that since our starting point is motivated from the application point of view, we adopt the following properties to propose the functional form of the integrand in Equation (18): directional dependency which is encoded in the first derivative of the underlying field and also inspired by the definition of crossing statistics \citep{ryden1988area}; intuitively, our suggestion belongs to the moment and cumulant definition of density field which is more reasonable compared to other complicated functions; taking into account other typical functions, namely $f(|\nabla \delta^{(r,s)}|^{n},\nabla^2 \delta^{(r,s)},  \nabla_{i} \nabla_{j}\delta^{(r,s)} \nabla^{i}\nabla^{j}\delta^{(r,s)}, ...)$ is in principle allowed but it turns out that the higher derivative the higher computational time consuming and even opens new room for the higher value of numerical uncertainty. Generally, the shear tensor $(\delta_{,ij}^{(r,s)})$ and its combination with the first derivative of the field which is represented by a generic definition of spectral parameters $\gamma_n\equiv \frac{\sigma^2_n}{\sigma_{n-1}\sigma_{n+1}}$ and characteristic radius of local extrema $(R_*\sim \sigma_1/\sigma_2)$ are relevant when we are dealing with the local extrema \citep{Bardeen:1985tr,2021MNRAS.503..815V}. As long as our purpose is focusing on directional dependency, we do not need to examine the extrema condition expressed by the second derivatives, consequently, a reasonable choice is adopting the first derivative of density field. Using the  probabilistic framework presented in the subsection
\ref{section:Probabilistic Framework1}, the expected values of $N_{cmd}^{(r,s)}$ for a  3-Dimensional
Gaussian density field is obtained as follows:

\begin{widetext}
\begin{eqnarray}
		\left\langle N_{cmd}^{(r,s)}(\vartheta,i;n) \right\rangle_G =   2^{\frac{n-3}{2}} [1+ \cos(n\pi)] \Gamma \left(\frac{n+1}{2}\right)
		\times\frac{\left(\sigma_{1i}^{(r,s)}\right)^n}{\pi\sigma_{0}^{(r,s)}}\;e^{-\vartheta^2/2}
	\label{eq:Nccr}
\end{eqnarray}
\end{widetext}
where $\Gamma(:)$ is Gamma function. Equation  (\ref{eq:Nccr}), implies that only the even value of $n$  is survived in the Gaussian regime and all odd values of first derivative moments are identically zero. To mitigating the numerical error, the lowest power adopted for the RSD analysis in the context of $cmd$ measure, would be $n=2$, throughout this paper.

\subsection{Perturbative Formalism}
In the previous subsection, we introduced our new measure, and in principle according to Equation   (\ref{eq:expected value}), we can derive the perturbative form of the $N_{cmd}^{(r,s)}$ for 3-Dimensional density field in the mildly non-Gaussian regime.  To this end, we expand the Equation   (\ref{eq:expected value})
for a typical observable quantity up to $\mathcal{O}(\sigma_0^2)$ as:
\begin{eqnarray}
	\begin{split}
		\langle \mathcal{F} \rangle =& \langle \mathcal{F} \rangle_G+ \frac{\sigma_{0}}{3!}\sum_{\mu_1,\mu_2,\mu_3}
		\mathcal{K}^{(3)}_{\mu_1,\mu_2,\mu_3}\langle\partial {\mathcal{F}}/\partial {\mathcal{A}}_{\mu_1}\partial {\mathcal{A}}_{\mu_2}\partial {\mathcal{A}}_{\mu_3}\rangle_G\\
		&+ \mathcal{O}(\sigma_{0}^2)
	\end{split}
	\label{eq:expansion}
\end{eqnarray}
Subsequently, the weakly non-Gaussian form of $N_{cr}^{(r,s)}$ and $N_{cmd}^{(r,s)}$, up to the $\mathcal{O}(\sigma_0^2)$, becomes:
\begin{widetext}
\begin{eqnarray}\label{eq:crnon}
	\left\langle N_{cr}^{(r,s)}(\vartheta,i)\right\rangle_{NG}  =
	\frac{\sigma_{1i}^{(r,s)}}{\pi\;\sigma_{0}^{(r,s)}}\;e^{-\vartheta^2/2} \;
	\bigg\{1\; +\;
	\bigg [ \frac{1}{6} \;S_0^{(r,s)} H_3(\vartheta) \; + \;\frac{1}{2}\frac{S_{1i}^{(r,s)}}{\sigma_0^{(r,s)}} H_1(\vartheta) \bigg ]\sigma_0^{(r,s)} \;+\; \mathcal{O}\bigg ((\sigma_0^{(r,s)})^2\bigg ) \, \bigg\},
	\end{eqnarray}
	\begin{eqnarray}\label{eq:cmdnon}
		\left\langle N_{cmd}^{(r,s)}(\vartheta,i)\right\rangle_{NG}  =
		\frac{\left(\sigma_{1i}^{(r,s)}\right)^2}{\sqrt{2\pi}\;\sigma_{0}^{(r,s)}}\;e^{-\vartheta^2/2} \;
		\bigg\{1\; +\;
		\bigg [ \frac{1}{6} \;S_0^{(r,s)} H_3(\vartheta) \; + \;S_{1i}^{(r,s)} H_1(\vartheta) \bigg ]\sigma_0^{(r,s)} \;+\; \mathcal{O}\bigg ((\sigma_0^{(r,s)})^2\bigg ) \, \bigg\},
	\end{eqnarray}
\end{widetext}
where $H_n(\vartheta)$ represent the probabilists' Hermite polynomials  and we have used following definitions:
\begin{eqnarray}
	S_0^{(r,s)}  \equiv \frac{\left\langle\;\left(\delta^{(r,s)}\right)^3\; \right\rangle_c\;}{\left(\sigma_{0}^{(r,s)}\right)^4},
\end{eqnarray}
\begin{eqnarray}
	S_{1i}^{(r,s)}  \equiv \;\frac{\left\langle\;\delta^{(r,s)} \left(\delta_{,i}^{(r,s)}\right)^2\; \right\rangle_c}{\sigma_0^{(r,s)}\;\left(\sigma_{1i}^{(r,s)}\right)^2},
\end{eqnarray}
Having Equations (\ref{eq:crnon}) and (\ref{eq:cmdnon}), we can predict the $\langle N_{cr}\rangle $ and $\langle N_{cmd}\rangle $ for a given field considering the corresponding spectral indices, respectively. We should notice that for implementation on the normalized density contrast field which is usually adopted by following transformation: 
	\begin{equation}
	\delta \longrightarrow \delta^{\prime} = \frac{\delta}{\sigma_0} \;\;,
	\end{equation}
	 the $n$th conditional moment of first derivative for $\delta^{\prime}$ with respect to that of for $\delta$, becomes:
	 \begin{eqnarray}\label{eq:transformation}
	 \left\langle N_{cmd}^{(r,s)}\right\rangle\Bigg|_{\delta^{\prime}} =\frac{\left\langle N_{cmd}^{(r,s)} \right\rangle\Bigg|_{\delta}}{\sigma_0^{(n-1)}}
	 \end{eqnarray}
	while the $N_{cr}$ and MVs are invariant against mentioned transformation.

\section{Implementation on the Redshift space}\label{sec:IRSD}
In this section, we consider the linear Kaiser and FoG effects as the sources of anisotropy in the density field and evaluate the imprint of these effects on our introduced measures in the previous section. Throughout this paper, we use the plane-parallel approximation and consider the $z$-axis of the Cartesian coordinate as the line of sight direction, without losing generality. In this approximation, there is no statistical difference between the directions perpendicular to $(\hat{{z}})$ (e.g. $\hat{{x}}$ and $\hat{{y}}$), and we use the notation $\hat{\mathcal{I}}$ to show these directions.

\subsection{The $cmd$ and $cr$ measures in the Linear Kaiser limit}
\begin{figure}
			\centering
	\includegraphics[width=0.8\columnwidth]{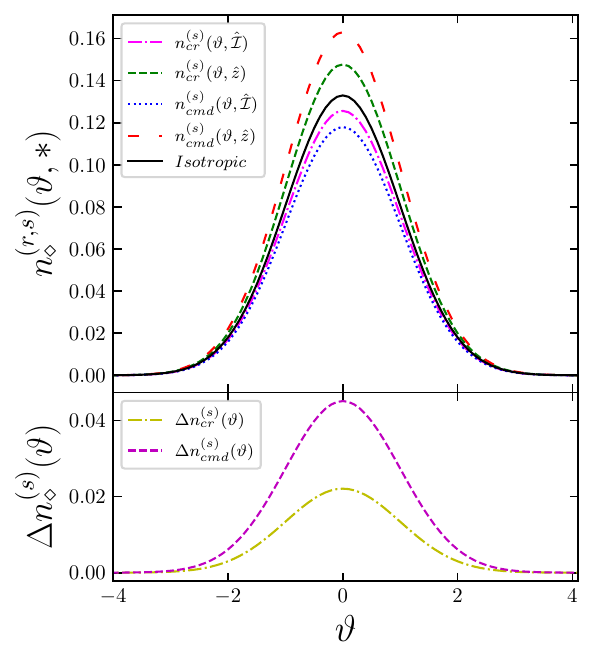}
	\caption{The theoretical prediction for the normalized $cr$ and $cmd$ measures as a function of threshold for $\beta_{\rm fiducial}=0.48$ as a fiducial value in the linear Kaiser limit. The  ``*'' symbol is replaced by $\hat{\mathcal{I}}$ and $\hat{z}$ for perpendicular and along to the line of sight, respectively. The $\diamond$ symbol is reserved for $cr$ and $cmd$ statistics.  The black solid line is for $\beta=0$ which shows the isotropic limit. The lower panel depicts the difference $\Delta n_{\diamond}^{(s)}(\vartheta)\equiv n_{\diamond}^{(s)}(\vartheta,\hat{z})-n_{\diamond}^{(s)}(\vartheta,\hat{\mathcal{I}})$ demonstrating that $\Delta n_{cmd}^{(s)}$ is higher than $\Delta n_{cr}^{(s)}$.   }
	\label{fig:example_figure}
\end{figure}

In the linear Kaiser limit, Equation   (\ref{eq:power spectrum}) reduces to \citep{Kaiser1987}:
\begin{eqnarray}
	P^{(s)}(\boldsymbol{k}) = (1+\beta\mu^2)^2P^{(r)}(\boldsymbol{k})
	\label{eq:linear power spectrum}
\end{eqnarray}
Using Equations    (\ref{eq:spectral0}), (\ref{eq:spectral00})  and   (\ref{eq:linear power spectrum}), one can obtain:
\begin{eqnarray}
	\left(\sigma_{0}^{(s)}\right)^2 &=& C_0\left(\sigma_{0}^{(r)}\right)^2,\nonumber\\
	\left(\sigma_{1}^{(s)}\right)^2 &=& C_0\left(\sigma_{1}^{(r)}\right)^2,
\end{eqnarray}
and
\begin{eqnarray}
	\left(\sigma_{{\mathcal{I}}}^{(s)}\right)^2 &=&  \left(\frac{C_0-C_1}{2}\right)\left(\sigma_{1}^{(r)}\right)^2,\nonumber\\ \left(\sigma_{1{z}}^{(s)}\right)^2 &=& C_1\left(\sigma_{1}^{(r)}\right)^2,
\end{eqnarray}
where
\begin{eqnarray}
	C_n(\beta) \equiv \frac12\int_{-1}^{1}\mu^{2n}\;(1+\beta\mu^2)^2\; d\mu\;,
	\label{eq:C_n}
\end{eqnarray}
Consequently, in the  linear Kaiser limit, the $cr$ and $cmd$ statistics for a 3-Dimensional Gaussian field in redshift space for $\hat{z}$ and $\hat{\mathcal{I}}$ directions become:
\begin{eqnarray}
	N_{cr}^{(s)}(\vartheta,\hat{\mathcal{I}}) = \frac{\sigma_{1}^{(r)}}{\sqrt{2}\pi\sigma_{0}^{(r)}}\;\sqrt{1-\frac{C_1}{C_0}}\;e^{-\vartheta^2/2},
\end{eqnarray}
\begin{eqnarray}
	N_{cr}^{(s)}(\vartheta,\hat{z}) = \frac{\sigma_{1}^{(r)}}{\pi\sigma_{0}^{(r)}}\;\sqrt{\frac{C_1}{C_0}}\;e^{-\vartheta^2/2},
\end{eqnarray}
Also 
\begin{eqnarray}
	N_{cmd}^{(s)}(\vartheta,\hat{\mathcal{I}}) = \frac{\left(\sigma_{1}^{(r)}\right)^2}{\sqrt{2\pi}\;\sigma_{0}^{(r)}}\;{\frac{C_0 - C_1}{2\;\sqrt{C_0}}}\;e^{-\vartheta^2/2},
\end{eqnarray}
\begin{eqnarray}
	N_{cmd}^{(s)}(\vartheta,\hat{z}) = \frac{\left(\sigma_{1}^{(r)}\right)^2}{\sqrt{2\pi}\;\sigma_{0}^{(r)}}\;{\frac{C_1}{\;\sqrt{C_0}}}\;e^{-\vartheta^2/2},
\end{eqnarray}
Note that the r.h.s of the above Equations have been expressed in terms of the real space spectral indices. For further analysis, we define the following normalized quantity for direction $i$:
\begin{eqnarray}
	\begin{split}
		&n_{\diamond}^{(r,s)}(\vartheta,i) \equiv\\ 
		& \frac{N_{\diamond}^{(r,s)}(\vartheta,i)}{\int_{-\infty}^{\infty}d\vartheta\left[N_{\diamond}^{(r,s)}(\vartheta,i)+N_{\diamond}^{(r,s)}(\vartheta,j)+N_{\diamond}^{(r,s)}(\vartheta,k)\right]},
	\end{split}\label{eq:normal_cr}
\end{eqnarray}
where $\{i,j,k\}\in[\hat{x},\hat{y}, \hat{z}]$ and $i\ne j\ne k$. The $\diamond$ is replaced by $cr$ and $cmd$. Interestingly,  the isotropic Gaussian limit of  Equation   (\ref{eq:normal_cr}) reduces to:
\begin{eqnarray}
	n_{cr}^{(r)}(\vartheta) = n_{cmd}^{(r)}(\vartheta)= \frac{1}{3\sqrt{\;2\pi}}\;e^{-\vartheta^2/2}
	\label{eq:isotropic limit}
\end{eqnarray}
Equation   (\ref{eq:isotropic limit})  reveals that in the isotropic Gaussian limit, normalized quantities are independent from the spectral indices and therefore the properties of the power spectrum. For a given field, any departure from Equation   (\ref{eq:isotropic limit})  can be considered as the signature of anisotropy and/or non-Gaussianity. While for the redshift space, the normalized quantities can be derived as:
\begin{eqnarray}
	n_{cr}^{(s)}(\vartheta,\hat{\mathcal{I}}) = \frac{1}{\sqrt{2\pi}}\;\frac{1}{2+\sqrt{\frac{2C_1}{C_0-C_1}}}\;e^{-\vartheta^2/2},
	\label{eq:n1}
\end{eqnarray}
\begin{eqnarray}
	n_{cr}^{(s)}(\vartheta,\hat{z}) = \frac{1}{\sqrt{2\pi}}\;\frac{1}{1+\sqrt{2}\sqrt{\frac{C_0}{C_1}-1}}\;e^{-\vartheta^2/2},
\end{eqnarray}
\begin{eqnarray}\label{eq:n44}
	n_{cmd}^{(s)}(\vartheta,\hat{\mathcal{I}}) = \frac{1}{2\sqrt{2\pi}}\;\left(1-\frac{C_1}{C_0}\right)\;e^{-\vartheta^2/2},
\end{eqnarray}
\begin{eqnarray}
	n_{cmd}^{(s)}(\vartheta,\hat{z}) = \frac{1}{\sqrt{2\pi}}\;\left(\frac{C_1}{C_0}\right)\;e^{-\vartheta^2/2},
	\label{eq:n4}
\end{eqnarray}
They have no explicit dependencies on the spectral indices. Thus, for the Gaussian limit, the normalized quantities only depend on the threshold, $\vartheta$, and redshift space parameter, $\beta$, through the $C_0$ and $C_1$. From Equation   (\ref{eq:C_n}), we find:
\begin{eqnarray}
	C_0(\beta) = 1+\frac{2\beta}{3}+\frac{\beta^2}{5} \;,\hspace{0.4cm} C_1(\beta) = \frac{1}{3}+\frac{2\beta}{5}+\frac{\beta^2}{7}\;.
\end{eqnarray}
In the limit $\beta \rightarrow 0$, Equations (\ref{eq:n1})-(\ref{eq:n4}) get the isotropic limit presented in Equation   (\ref{eq:isotropic limit}). In such a limit, the normalized $cr$ and $cmd$ measures are similar. Therefore, for $\beta\ne0$, the among of deviation from the isotropic limit can be considered as a signature for determining the sensitivity of $cr$ and $cmd$ statistics to RSD. In Fig. \ref{fig:example_figure}, we plot the analytical predictions of the normalized $cr$ and $cmd$ quantities as a function of threshold, $\vartheta$, for a typical anisotropic Gaussian matter field in redshift space in the presence of the linear Kaiser effect adopted by  $\beta= 0.48$ as a fiducial value. The black solid line illustrates the normalized  $cr$ and $cmd$ for isotropic limit (Equation   (\ref{eq:isotropic limit})). The green dashed line corresponds to $n_{cr}$ for a line of sight direction, while the purple dashed-dotted line is perpendicular to the line of sight direction. The linear Kaiser effect squeezes the isodensity contours along the line of sight. As a result, according to the definition of $cr$ statistics, we expect that the value of $n_{cr}$ is higher for the line of sight direction compared to $\hat{\mathcal{I}}$ directions. Noticing the analytical form of $cmd$ criterion (Equations    (\ref{eq:n44}) and (\ref{eq:n4})), the mentioned imprint of linear Kaiser effect would be magnified leading to make a robust measure compared to the common crossing statistics. Subsequently, the difference between $n_{cmd}(\hat{\mathcal{I}})$ (blue dotted line) and $n_{cmd}(\hat{z})$ (red loosely dashed line) is higher than the corresponding value in the context of $cr$ measure for as fixed value of $\beta$. The lower panel of Fig.  \ref{fig:example_figure} indicates the difference between $\Delta n_{\diamond}^{(s)}(\vartheta)\equiv n_{\diamond}^{(s)}(\vartheta,\hat{z})-n_{\diamond}^{(s)}(\vartheta,\hat{\mathcal{I}})$.

To make more complete our discussion regarding the capability of $cr$ and $cmd$ measures to put the observational constraint on the RSD parameter, we follow the approach carried on by  \citet{appleby2019ensemble} in the context of Minkowski tensors. We introduce following quantities by means of $cr$ and $cmd$ criteria as:
\begin{eqnarray}
	\varTheta_{cr}(\vartheta)&\equiv& \frac{N_{cr}^{(s)}(\vartheta,\hat{\mathcal{I}})}{N_{cr}^{(s)}(\vartheta,\hat{z})},\nonumber\\ \varTheta_{cmd}(\vartheta)&\equiv& \frac{N_{cmd}^{(s)}(\vartheta,\hat{\mathcal{I}})}{N_{cmd}^{(s)}(\vartheta,\hat{z})}.
\end{eqnarray}
It turns out that for the Gaussian and linear Kaiser limit, we have:
\begin{eqnarray}\label{eq:42}
	\varTheta_{cr}&=& \sqrt{\frac{C_0(\beta)-C_1(\beta)}{2 C_1(\beta)}},\nonumber\\ \varTheta_{cmd}&=& \frac{C_0(\beta)-C_1(\beta)}{2 C_1(\beta)}.
\end{eqnarray}
\begin{figure}[t]
			\centering
	\includegraphics[width=0.8\columnwidth]{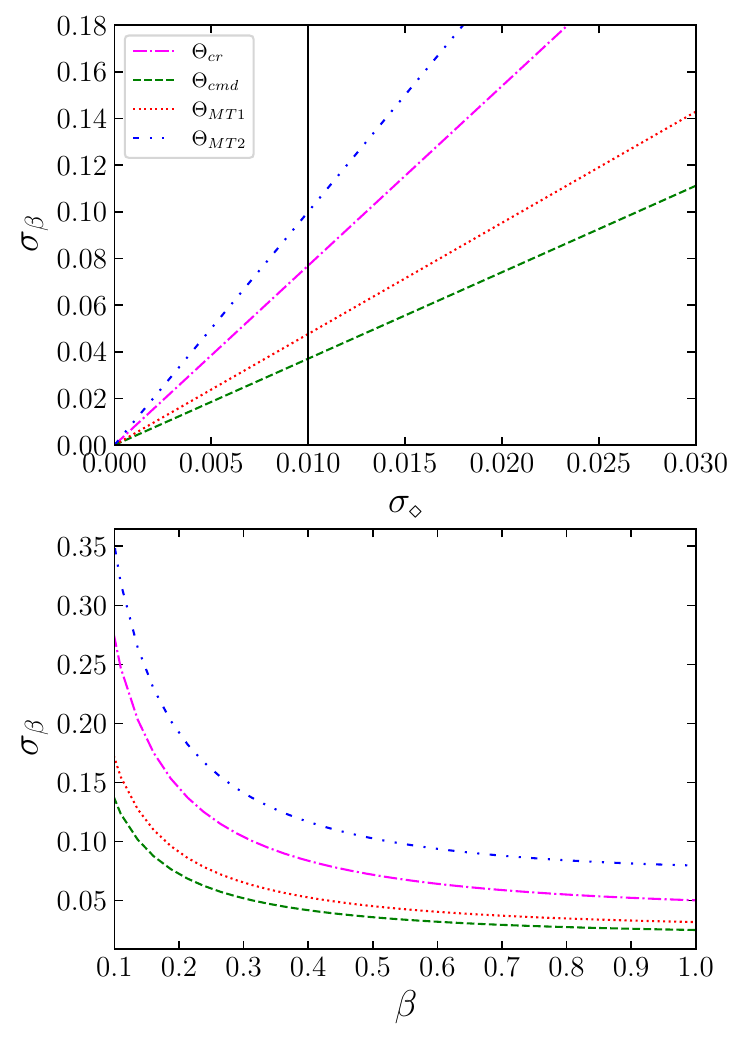}
	\caption{Upper panel: Error propagator on the $\beta$ for various criteria discussed in the text. Supposing  one percent relative error produced for $\sigma_{\diamond}$ in observation causes lower relative error on redshift space parameter by $cmd$ measure compared to other criteria examined in this paper. Lower panel: The $\beta$-dependancy of relative error ($\sigma_{\beta}$) for various statistics. We considered $\sigma_{\diamond}=0.01$.}
	\label{Fisher}
\end{figure}
We use the notation  $\varTheta_{\diamond}$ with $ \diamond \in \{cr,cmd,MT1,MT2\}$. Here the $MT1$ and $MT2$ are associated with the type one and two of rank-2 Minkowski tensors as defined by  \cite{appleby2019ensemble}. Based on the theoretical predictions of $\varTheta_{\diamond}$, we can determine the level of accuracy accordingly, we can constrain the value of parameter $\beta$. This accuracy depends on the statistical uncertainty associated with $\varTheta_{\diamond}$, which can be evaluated using the Fisher forecast approach. We rely on  the posterior probability function, $\mathcal{P}_{\diamond}(\beta|\Theta_{\diamond})$, as:
\begin{eqnarray}
	\begin{split}
		\mathcal{P}_{\diamond}(\beta|\varTheta_{\diamond})&=\langle\delta_{D}(\beta-\Phi^{-1}_{\diamond}(\varTheta_{\diamond}))\rangle\\
		&=\int d\varTheta'_{\diamond}\mathcal{P}(\varTheta'_{\diamond})\delta_D(\varTheta_{\diamond}-\varTheta'_{\diamond})\left|\mathcal{J}\right|_{\varTheta'_{\diamond}=\Phi_{\diamond}(\beta)}
	\end{split}
\end{eqnarray}
here $\mathcal{J}$ is the Jacobian computed for $\varTheta'_{\diamond}=\Phi_{\diamond}(\beta)$ and $\Phi_{\diamond}(\beta)$ reads by Equation   (\ref{eq:42}). Finally, according to a given confidence interval (C.L.), the error-bar on $\beta$ is given by:
\begin{eqnarray}
	{\rm C.L.}=\int_{\beta_{\rm fiducial}-\sigma^{(-)}_{\beta}}^{\beta_{\rm fiducial}+\sigma^{(+)}_{\beta}}d\beta\; \mathcal{P}_{\diamond}(\beta|\varTheta_{\diamond})
\end{eqnarray}
or equivalently, based on the error propagation formalism up to the first order, we obtain the relative error on redshift space parameter as:
\begin{eqnarray}
	\sigma^2_{\beta} = \left( \frac{\partial \ln\varTheta_{\diamond}}{\partial \ln \beta} \right)^{-2} \sigma^2_{\diamond}
\end{eqnarray}
where $\sigma_{\beta}$ and $\sigma_{\diamond}$ represent the fractional uncertainties on $\beta$ and $\varTheta_{\diamond}$, respectively.
In the upper panel of Fig.   \ref{Fisher}, we plot $\sigma_{\beta}$ in terms of $\sigma_{\diamond}$ for $\beta_{\rm fiducial} = 0.48$ as the fiducial value in the Gaussian limit. We also consider $\sigma_{\diamond} = 0.01$ as the comparison base value which is shown by the black vertical solid line in this figure. As mentioned in previous research,  incorporating the one percent relative error in the statistical measures as already has been achieved by the current galaxy catalogs, yielding almost higher accuracy in the context of $cmd$ criterion compared to all other statistics including $cr$ and rank-2 Minkowski tensors. 
Since the functional form of $\mathcal {G}$ to establish {\it cmd} statistics compared to the common generalization of the MFs, particularly the rank-2 MTs causes to manipulate the presence of the field first derivative for different directions, namely, $(\sigma_{1x},\sigma_{1y}, \sigma_{1z})$, in the denominator of MT1 resulting in almost increasing the sensitivity of $cmd$. The $\beta$-dependency of sensitivity with respect to $\varTheta_{\diamond}$ (ratio quantity) for different measures is shown in the lower panel of Fig. \ref{Fisher}. The relative difference of $\sigma_\beta$ for $MT1$ respect to the $cmd$ measure demonstrates that utilizing $cmd$ yields almost $20\%$ improvement which is almost accepted to achieve high precision evaluation.

\begin{figure*}
	\centering
	\includegraphics[width=0.4\textwidth]{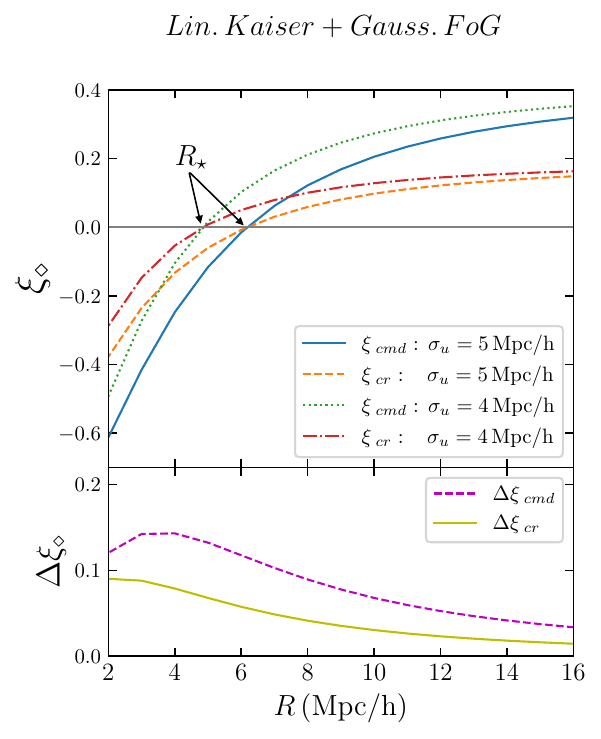}
	\includegraphics[width=0.4\textwidth]{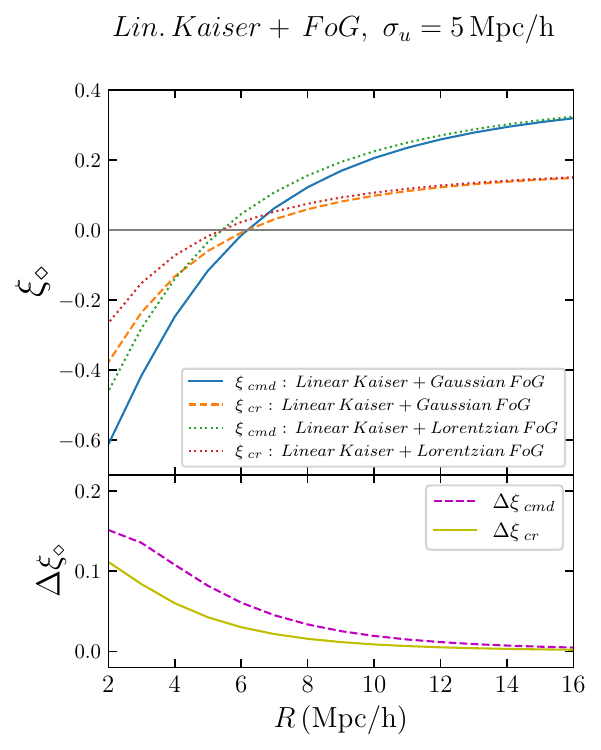}
	\caption{Left panel: $\xi_{\diamond}$ as a function of $R$ for the phenomenological Gaussian model of FoG and for $\sigma_u=4$ Mpc h$^{-1}$ and $\sigma_u=5$ Mpc h$^{-1}$. There is a trade off between the imprint of FoG and linear Kaiser effects for different smoothing scales due to their contradiction behaviors at small and large scales, respectively. At the so-called $R_{\star}$ whose value depends on cosmological parameters, the directional dependency of $cr$ and $cmd$ is negligible. The lower part of the left panel illustrates the difference of $\xi{\diamond}$ for both velocity dispersions. Right panel: the comparison between two phenomenological models for FoG, namely the Lorentzian and the Gaussian models, in the context of $cr$ and $cmd$ statistical measures. The corresponding lower panel depicts the difference between $\xi_{\diamond}$ for the Gaussian and the Lorentzian cases.  }
	\label{fig:FoG}
\end{figure*}

\subsection{Finger of God impact on the $cr$ and $cmd$ measures}
Thus far, we have applied $cr$ and $cmd$ statistics to the redshift space density field in the presence of the linear Kaiser effect. In this subsection, we take into account the FoG phenomenon in addition to the linear Kaiser effect as the anisotropy sources of the density field in redshift space and try to characterize their impacts on our statistical measures.

To elaborate on the FoG effect describing the elongation of the clusters along the line of sight on small scales,
there are several phenomenological models in the literature. Here we consider Gaussian \citep{Peacock1994} and Lorentzian \citep{Percival2004,Ballinger1996} FoG models which are respectively read off by the following Equations:
\begin{eqnarray}
	\tilde{O}^{\rm Gauss}_{\rm FoG}(k\mu,\sigma_u) = e^{-\frac{1}{2}\sigma_u^{2} k^2 \mu^2}
	\label{eq:Gauss}
\end{eqnarray}
and
\begin{eqnarray}
	\tilde{O}^{\rm Lorentz}_{\rm FoG}(k\mu,\sigma_u) = \frac{1}{1+\frac{1}{2}\sigma_u^{2} k^2 \mu^2} ,
	\label{eq:Lorentz}
\end{eqnarray}
where $\sigma_u$ is the one-dimensional velocity dispersion. More precisely, to manipulate the linear Kaiser and FoG effects together, the non-linear part of the Equation   (\ref{eq:operator}) can be replaced by the Equation   (\ref{eq:Gauss}) or Equation   (\ref{eq:Lorentz}). It is worth noting that the spectral indices given by Equations   (\ref{eq:spectral0}) and (\ref{eq:spectral00}) are modified by correction of power spectrum which is in principle constructed by plugging the Equation   (\ref{eq:Gauss}) or Equation   (\ref{eq:Lorentz}) in the Equation   (\ref{eq:power spectrum}).

To go further, we define $\xi_{\diamond} \equiv \varTheta^{-1}_{\diamond} - 1$ and  $\diamond \in \{cr,cmd\}$. The Gaussian limits of $\xi_{cr}$ and $\xi_{cmd}$ are obtained as:
\begin{eqnarray}\label{eq:xi1}
	\xi_{cr}= \frac{\sigma^{(s)}_{1z}}{\sigma^{(s)}_{1\mathcal{I}}}-1, \hspace{0.7cm} \xi_{cmd}= \left(\frac{\sigma^{(s)}_{1z}}{\sigma^{(s)}_{1\mathcal{I}}}\right)^2-1,
\end{eqnarray}
Therefore, in this case, $\xi_{\diamond}$ statistics depend on the following quantities including parameter $\beta$, FoG model,  one dimensional velocity dispersion, $\sigma_u$, smoothing kernel, $\tilde{W}$, smoothing scale, $R$, and power spectrum in the real space, $P^{(r)}(\boldsymbol{k})$. The $\xi_{\diamond}$ can be numerically computed for a desired cosmological field and supposing the Gaussian model, this can be considered as a new model-dependent observational measure to constrain the associated cosmological parameters.

To compare the FoG contribution with the linear Kaiser effect, we adopt a Gaussian smoothing kernel equates to $\tilde{W}(kR) = \exp(-(kR)^2/2)$. We also use CAMB software \citep{Lewis:1999bs} with the fiducial values $\{\,\Omega_{\Lambda} = 0.69179,\, \Omega_{c}h^2 = 0.11865,\, \Omega_{b}h^2 =0.022307,\, \Omega_{\nu}h^2 = 0.000638,\, h = 0.6778,\,\, n_s = 0.9672\, \}$ which is in agreement with the flat $\Lambda$CDM $Planck$ cosmological parameters to compute the matter power spectrum,  \citep{aghanim2020planck,aghanim2020planckvi}.
Adopting the linear bias  $b=1$, we obtain $\beta = 0.526$ for these cosmological parameters.

In the left panel of Fig.   \ref{fig:FoG}, we illustrate the $\xi_{\diamond}$ as a function of the smoothing scale when the phenomenological Gaussian  model is taken into account for the FoG effect. Scaling dependency of $\xi_{\diamond}$ is clearly due to the FoG and interestingly we obtain that at a smoothing scale denoted by $R_{\star}$, the $\xi_{\diamond}$ pierces the zero threshold. This means that at such a scale, the directional dependency of $cr$ and $cmd$ measures are diminished due to the competition between the linear Kaiser and FoG effects which behave on the contrary ways. In other words, the FoG and the linear Kaiser effects lead to the stretching and hardening of the iso-density contours along the line of sight, respectively. On a specific scale ($R_{\star}$), the linear Kaiser effect and the FoG effect cancel each other out, and $\xi_{cr}$ and $\xi_{cmd}$ reach zero. In scales smaller than $R_{\star}$, the FoG effect is dominant, and both $\xi_{cr}$ and $\xi_{cmd}$ have negative values, but for scales larger than $R_{\star}$, the linear Kaiser effect becomes significant and both $\xi_{cr}$ and $\xi_{cmd}$ take positive values. In addition, by increasing the velocity dispersion, the dominant range of FoG grows, which is in agreement with the analytical modeling of FoG. In the lower part of left panel, we plot the $\Delta \xi_{\diamond}\equiv\xi_{\diamond}(\sigma_u=5\;{\rm Mpc h}^{-1})-\xi_{\diamond}(\sigma_u=4\;{\rm Mpc h}^{-1})$ and our results confirm that $\xi_{cmd}$ is higher than $\xi_{cr}$ for the two fixed values of $\sigma_u$.

In the right panel of Fig. \ref{fig:FoG}, the $\xi_{\diamond}$ for the Gaussian and Lorentzian models of FoG are compared. The higher value of smoothing scale leads to diminishing the contribution of higher $k$ resulting in the two mentioned models converge to each other. We must point out that, the sensitivity of $cr$ measure is almost less than $cmd$ criterion to non-linearity in redshift space. Therefore, the capability of $cmd$ in distinguishing different FoG models is higher than $cr$ statistics. To make more sense, we also compute  $\Delta \xi_{\diamond}\equiv\xi_{\diamond}(Lorentzian\; FoG)-\xi_{\diamond}(Gaussian\;FoG)$ and the bottom part of left panel shows that $\xi_{cmd}$ for small smoothing scale has higher dependency on the model of FoG than the $\xi_{cr}$. 

As indicated by the right panel of Fig. \ref{fig:FoG}, for small smoothing scale the $\xi_{cmd}$ has more $R$-dependency leading to have more dependency to the scale dependent bias, while the $\xi_{cr}$ has almost smaller amplitude and it shows weak $R$-dependency causing to have less dependency to scale dependent bias. This means that for the cosmological inferences from the linear regime, utilizing the $cr$ statistics reveals robust pipeline. Meanwhile, to put almost stringent constraints on the cosmological parameters and to examine the  peculiar velocity field \citep{2022PhRvD.105j3028J}, the $cmd$ measure may give promising results.
	\begin{figure}
		\centering
		\includegraphics[width=0.4\textwidth]{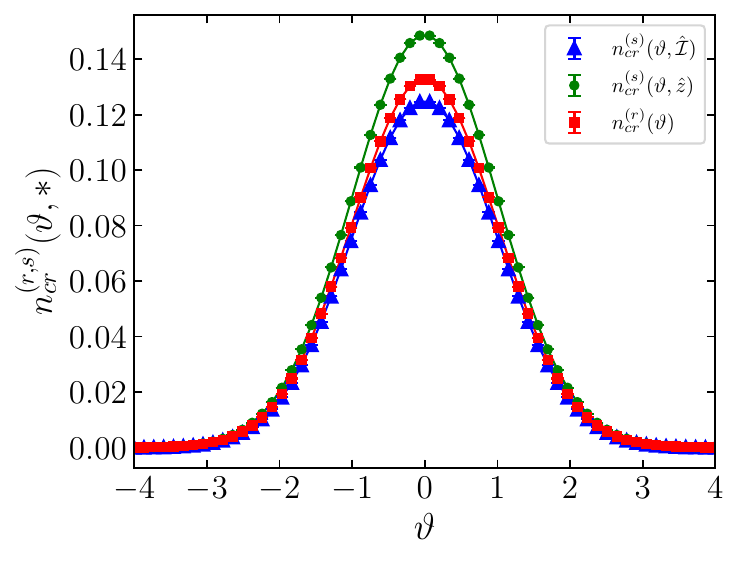}
		\hfill
		\includegraphics[width=0.4\textwidth]{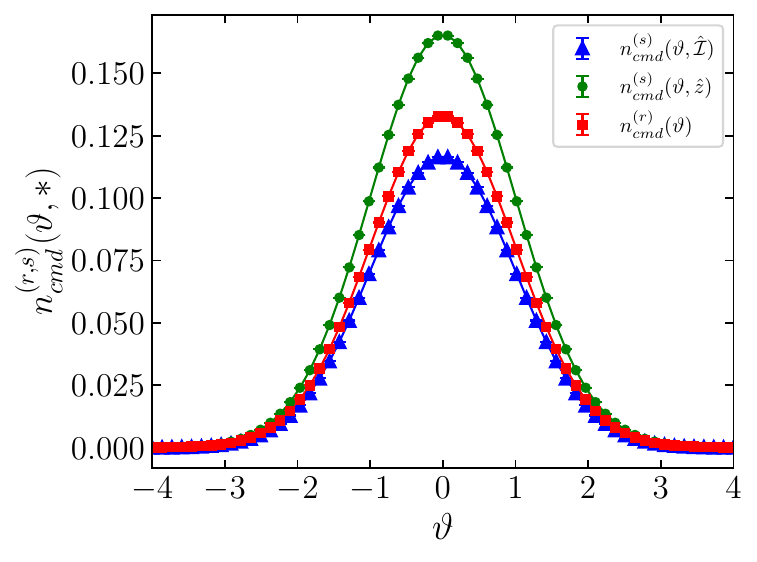}
		\caption{Upper panel: The crossing statistics as a function of $\vartheta$ for theoretical predictions (solid lines) and corresponding numerical reconstructions (symbols). Lower panel: $cmd$ measure versus threshold. The solid lines correspond to theoretical predictions, while the symbols indicate the results given by numerical simulations. Here we took $R=20$ Mpc h$^{-1}$. }
		\label{fig:wide_subplot}
	\end{figure}	

\begin{figure}
			\centering
	\includegraphics[width=0.8\columnwidth]{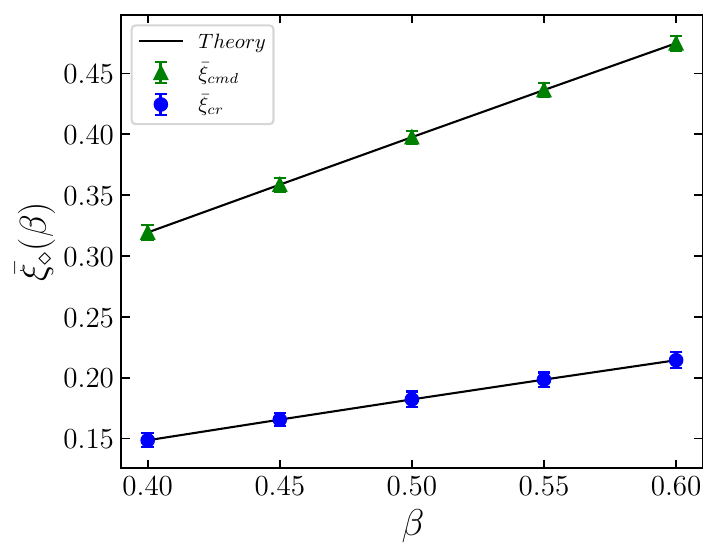}
	\caption{The $\bar{\xi}_{\diamond}$ as a function of $\beta$. The filled  circle symbols correspond to the numerical analysis of $cr$, while the filled triangles represent the numerical results for the $cmd$. The solid lines indicate the corresponding theoretical predictions. Here we took $R=20$ Mpc h$^{-1}$. }\label{fig:xibar}
\end{figure}
\begin{figure*}
	\includegraphics[width=2\columnwidth,height=0.32\textheight]{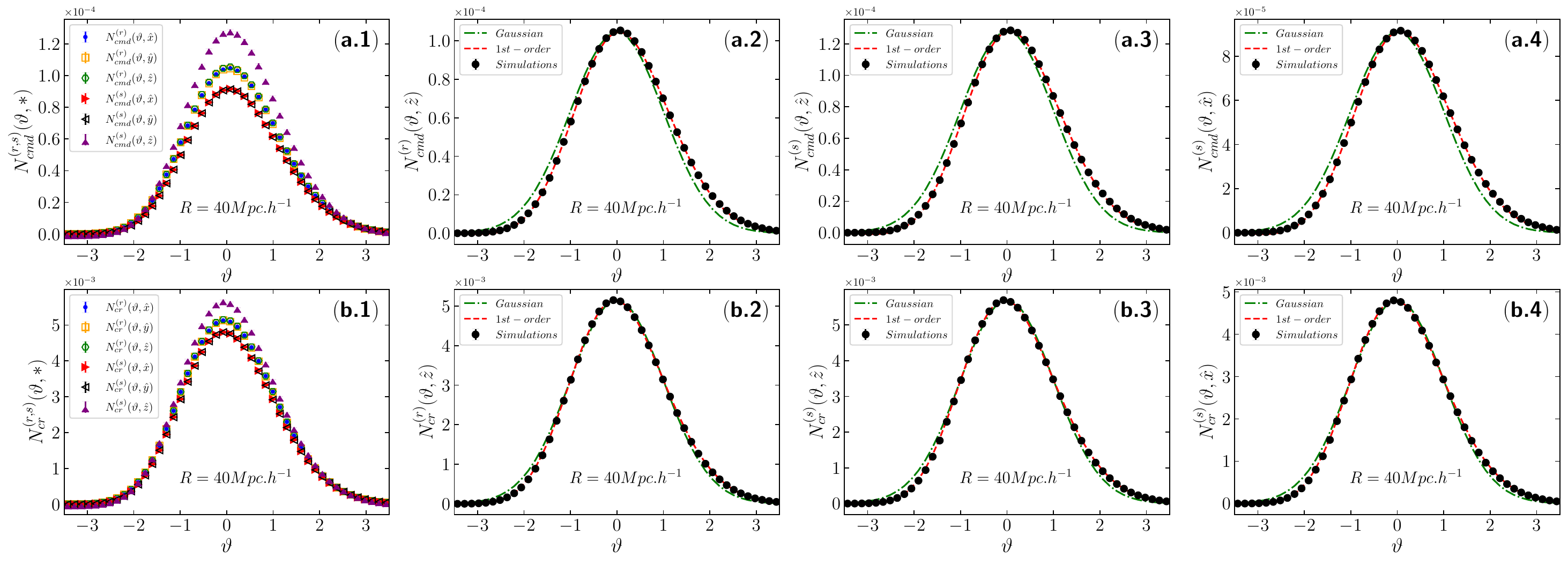}
	\caption{The $N^{(r,s)}_{(cmd)}$ [Mpc h$^{-1}$]$^{-2}$ and $N^{(r,s)}_{(cr)}$ [Mpc h$^{-1}$]$^{-1/2}$ versus threshold for the Quijote simulations. Panel (a.1): The expectation value of conditional moment of the first derivative in real and redshift spaces for both $\hat{\mathcal{I}}\in[\hat{x},\hat{y}]$ and $\hat{z}$ directions adopting $R=40$ Mpc h$^{-1}$.  Panel (a.2): The $N^{(r)}_{cmd}(\hat{z})$ for the Gaussian prediction considering corresponding spectral indices (green dashed-dot line) while the red dashed line indicates the theoretical non-Gaussian prediction for $cmd$ (Equation (\ref{eq:cmdnon})). The filled black circle symbols correspond to the numerical analysis including their $1\sigma$ level of confidence. The panels (a.3) and (a.4) are the same as the panel (a.2) just for redshift space in $\hat{z}$ and $\hat{x}$ directions, respectively. The lower panels are the same as the upper panels just for $N^{(r,s)}_{cr}$. }\label{fig:quij11}
\end{figure*}
\section{Application on Mock data}\label{sec:mockdata}
In this section, we are going to numerically extract the $cr$ and $cmd$  statistical measures  for  simulated anisotropic density field and compare our results with the theoretical predictions obtained in previous section. Two following approaches are considered: at first, according to the computed matter power spectrum consistent with flat $\Lambda$CDM model, we simulate Gaussian random field. Secondly, we will rely on the N-body simulations known as Quijote simulations \citep{Quijote_sims}.

\subsection{Gaussian synthetic field}

We consider the linear Kaiser effect as a source of anisotropy and therefore generate the anisotropic Gaussian field. To this end, using the linear power spectrum of matter determined by CAMB, we generate an isotropic Gaussian density field, $\delta^{(r)}$, sampled on a cubical lattice with the total volume size $V = (1{\rm Gpc\; h^{-1}})^3$ which consist of $N_{pix} = 512^3 $ pixels. Applying the Fourier transform on the simulated isotropic density field, we construct an anisotropic field according to the following transformation:
\begin{eqnarray}
	\tilde{\delta}^{(s)}(\boldsymbol{k}) = \left(1+\beta\frac{k_z^2}{k^2}\right)\tilde{\delta}^{(r)}(\boldsymbol{k}),
\end{eqnarray}
and therefore, we obtain the redshift space density field in Fourier space. Then we smooth $\tilde{\delta}^{(s)}$ by a Gaussian kernel with scale $R = 20$ Mpc h$^{-1}$. We generate $N_{sim} = 100$ realizations of Gaussian isotropic and anisotropic fields, and then apply the mentioned numerical methods to the simulated density fields to obtain the $cr$ and $cmd$ measures as a function of threshold.  For each realization, we extract these statistics in threshold range, $\vartheta\in[-4.0,4.0]$, 
and then we do the ensemble average. Increasing the number of realizations had no significant effect on our ensemble average.   
Fig.   \ref{fig:wide_subplot} presents the $cr$ and $cmd$ as a function of $\vartheta$ for perpendicular and along to line of sight in redshift space and for real space. The upper panel corresponds to $cr$, while the lower panel shows the $cmd$. The solid lines are associated with theoretical predictions and the symbols are for corresponding numerical results. The error bar represents the $1\sigma$ level of confidence demonstrating the good consistency between the numerical and theoretical results. However, due to the presence of the first derivative of the smoothed field in $cmd$ measure, we expect to obtain an almost higher value of error-bar compared to $cr$.

In the rest of this subsection, motivated by introducing a proper measure to put observational constraint on $\beta$, we define the following weighted summation to marginalize the effect of threshold bins:
\begin{eqnarray}\label{eq:xibar}
	\bar{\xi}_{\diamond} = \frac{\sum_{\vartheta_i=\vartheta_{\min}}^{\vartheta_{\max}}\omega_{\diamond}(\vartheta_i) \xi_{\diamond}(\vartheta_i)}{\sum_{\vartheta_i=\vartheta_{\min}}^{\vartheta_{\max}}\omega_{\diamond}(\vartheta_i)},
\end{eqnarray}
where the weight is defined by means of statistical error as $\omega_{\diamond}(\vartheta_i)\equiv \sigma^{-1}_{\xi_{\diamond}}(\vartheta_i)$.
In Fig.   \ref{fig:xibar}, we plot the $\bar{\xi}_{\diamond}$ statistics as a function of $\beta$ extracted from theoretical (solid line) and computational (symbols) approaches.  The higher slope for $cmd$ measure concerning the $cr$ statistics versus $\beta$ reveals more robustness in discriminating different values of $\beta$.

\begin{figure*}
	\includegraphics[width=2\columnwidth,height=0.32\textheight]{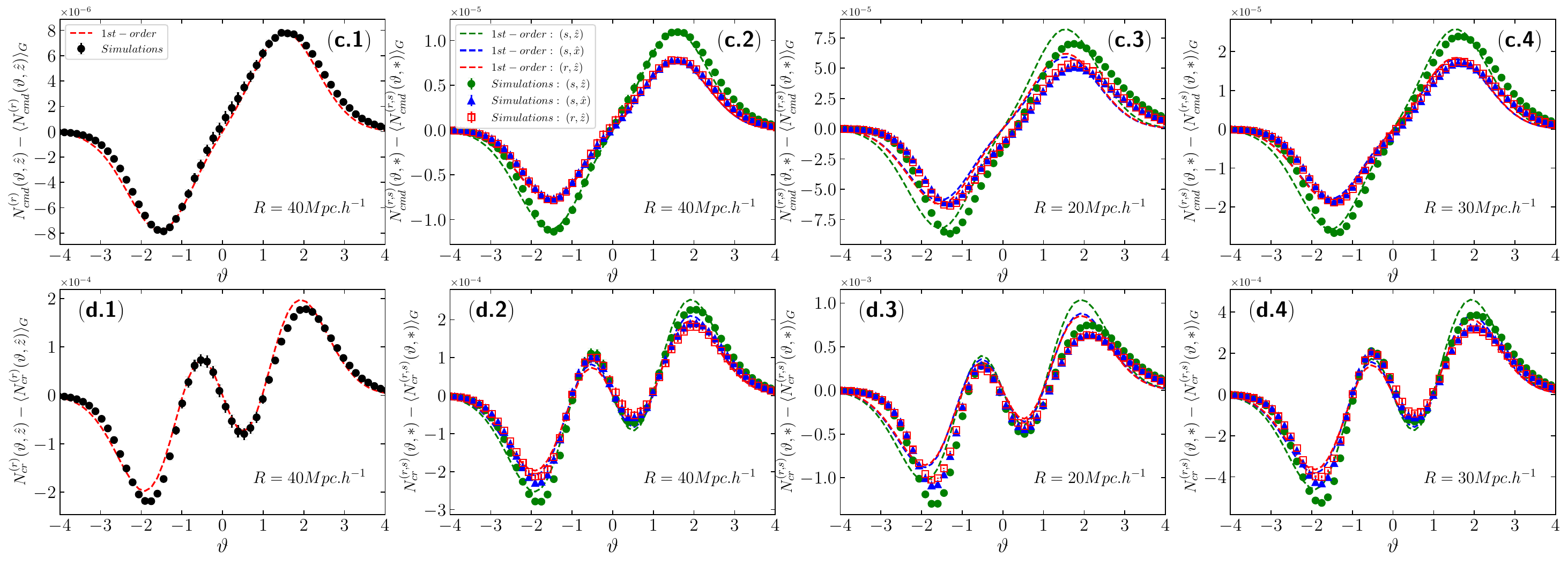}
	\caption{The $\left(N^{(r,s)}_{(cmd)}-\langle N^{(r,s)}_{(cmd)}\rangle_G\right)$ [Mpc h$^{-1}$]$^{-2}$ and $\left(N^{(r,s)}_{cr}-\langle N^{(r,s)}_{cr}\rangle_G\right)$ [Mpc h$^{-1}$]$^{-1/2}$  versus threshold for Quijote simulations. The panel (c.1) shows the difference between $N^{(r)}_{cmd}(\vartheta,\hat{z})$ and the value corresponds to the theoretical Gaussian prediction for $R=40$ Mpc h$^{-1}$. The red dashed solid line is for the theoretical prediction for deviation from Gaussian field up to the second order expansion, while the filled black circle symbols are for numerical deviation from  Gaussian theory. The same quantities just for redshift space are plotted in the panels (c.2), (c.3), and (c.4) for corresponding smoothing scales $R=40, 20, 30$ Mpc h$^{-1}$, respectively. Notice that, for the redshift space, the results for the $\hat{x}$ direction has been also given. The lower panels are the same as the upper panels just for the $cr$ measure.  }\label{fig:quij12}
\end{figure*}

\subsection{N-body simulations}
To examine the non-Gaussian impact on the conditional moments of derivative, we use the three-dimensional large scale structure made by publicly available N-body simulations from the Quijote complex \citep{Quijote_sims}. 
Each our ensemble extracted form the Quijote simulations has following properties: $N_{\rm particles}=512^3$, box size of $V=(1{\rm Gpc/h})^3$, the fiducial cosmological parameter are based on flat $\Lambda$CDM including $\Omega_m=0.3175$, $\Omega_b=0.049$, $h=0.6711$, $n_s=0.9624$ and $\sigma_8=0.834$ (see \citep{Quijote_sims} for more details). To construct the proper density field in applying our numerical pipeline, we use \texttt{Pylians} \citep{villaescusa2018pylians} and at redshift $z=0$, exploiting the routine cloud in cell (CIC) for mass assignment, the density field contrast, $\delta(\boldsymbol{r})$, would be retrieved. Finally, convolving the $\delta(\boldsymbol{r})$ with a Gaussian window function characterized by a smoothing scale, $R$, the matter density contrast is constructed in the real space, $\delta_R^{(r)}(\boldsymbol{r})$. To create the corresponding field in redshift space in plane-parallel approximation, $\delta_R^{(s)}(\boldsymbol{r})$, for each value of  $\boldsymbol{r}$, we use \texttt{Pylians}  which in principle considers the Equation   (\ref{eq:redshift position}). We also use a transformation to make density field with unit variance as $\delta \to \delta^{\prime}=\delta/\sigma_0$ \footnote{For this case the coefficient for $N_{cmd}^{(r,s)}$ gets following transformation:  $\frac{\left(\sigma_{1i}^{(r,s)}\right)^2}{\sigma_0^{(r,s)}}\to \left(\frac{\sigma_{1i}^{(r,s)}}{\sigma_0^{(r,s)}}\right)^2$.}.

The upper panels of Fig. \ref{fig:quij11} depicts the $N^{(r,s)}_{cmd}$ versus threshold for the Quijote simulations. The panel (a.1) corresponds to $N^{(r,s)}_{cmd}$ in real and redshift spaces for $\hat{\mathcal{I}}\in[\hat{x},\hat{y}]$ and $\hat{z}$ directions taking $R=40$ Mpc h$^{-1}$. As we expect, there are no significant deviations between various directions in real space, while in the redshift space, the $N^{(s)}_{cmd}(\vartheta,\hat{x})= N^{(s)}_{cmd}(\vartheta,\hat{y})\ne N^{(s)}_{cmd}(\vartheta,\hat{z})$. The panel (a.2) indicates the $N^{(r)}_{cmd}(\vartheta,\hat {z})$ for the Gaussian prediction (green dashed-dot line) while the red dashed line indicates the theoretical non-Gaussian prediction for $cmd$ (Equation (\ref{eq:cmdnon})). The filled black circle symbols correspond to the numerical analysis including their $1\sigma$ level of confidence. The numerical result is mildly skewed and it is tilted to the higher thresholds yielding the non-Gaussian behavior. The  panel (a.3) illustrates the $N^{(s)}_{cmd}(\vartheta,\hat{z})$ for Gaussian model (green dashed-dot line), non-Gaussian model up to the $\mathcal{O}(\sigma_0^2)$ (red dashed line) and filled circle symbols correspond to the numerical results for $R=40$ Mpc h$^{-1}$. The panel (a.4) shows the results for $N^{(s)}_{cmd}(\vartheta,\hat{x})$ which is perpendicular to the line of sight direction. The behavior of $N^{(r,s)}_{cr}$ for different situations are illustrated in the lower part of Fig. \ref{fig:quij11}. Our results verify that the $N_{cmd}$ is more capable to quantify the difference between various direction in redshift space. 

The difference between the numerical computation of $N^{(r,s)}_{(cmd,cr)}$ and the corresponding theoretical Gaussian prediction are depicted in Fig. \ref{fig:quij12}. The dashed lines are for the deviation of perturbative non-Gaussian theory concerning Gaussian prediction, while the symbols are the same quantities computed from simulations, numerically. In the  left panel (c.1), we depict the difference between the numerical results and theoretical Gaussian predictions in addition to the variation of the theoretical non-Gaussian model concerning the Gaussian form for real space. The green filled circle symbols, blue triangle symbols, and red rectangle symbols indicate the difference between $N^{(s)}_{cmd}(\vartheta,\hat{z})$, $N^{(s)}_{cmd}(\vartheta,\hat{x})$ and $N^{(r)}_{cmd}(\vartheta,\hat{z})$ computed numerically for N-body simulations and associated Gaussian models, respectively, in the panel (c.2). For this part, we consider the smoothed scale equates to $R=40$ Mpc h$^{-1}$. We display the same quantities as expressed for the  panel (c.2) but for $R=20$ Mpc h$^{-1}$ in the panel (c.3) and $R=30$ Mpc h$^{-1}$ in the panel (c.4). Our results confirm that the deviation from Gaussianity perpendicular to the line of sight directions in redshift space is almost the same as the real space. It has been shown that keeping the directional dependency in computing power spectrum and also in derived quantities causes to mitigate the degeneracy between RSD and non-linearity consequences \citep{2016MNRAS.457.1076J}. We also advocate that the separation of perpendicular to the line of sight analysis from the $\hat z$ direction in the redshift space can reduce the RSD impact on the cosmological inferences such as non-Gaussianity.  It is worth noting that, to compute the corresponding theoretical results, we adopt the spectral indices numerically from simulations. The inconsistency between theory and numerical results extracted from N-body simulations is justified due to the reason that for the lower value of the smoothing scale, the $\sigma_0$ gets higher value for the lower value of $R$, consequently, to obtain more precise consistency, we have to take into account the higher terms in perturbative formula according to Equation (\ref{eq:expansion}) to achieve more accurate formula for $\left\langle N_{cmd}^{(r,s)}(\vartheta,i)\right\rangle_{NG}$ (Equation (\ref{eq:cmdnon})). The behavior of $N^{(r,s)}_{cr}-\langle N^{(r,s)}_{cr}\rangle_G$  for different situations are illustrated in the lower part of Fig. \ref{fig:quij12}. Our results demonstrate that to have more consistent results from numerical analysis and theoretical prediction for $cr$ measure, we need to take into account higher order terms beyond 1st-order. In addition, for some lower thresholds, the results for the $hat{x}$ in the redshift space deviates from that of in real space in the context of $N_{cr}$  confirming that to mitigate the RSD non-Gaussianity imprint, the $\vartheta\gtrsim 0$ should be taken while this limitation almost does not exist for the $N_{cmd}$ statistics.

\subsection{Fisher Forecasts }
In order to present a quantitative description regarding the capability of various criteria explained before, particularly $cmd$ and $cr$ to put constraint on relevant cosmological parameters,  we compute the Fisher matrix in this subsection (see‌ e.g. \citep{2011IJMPD..20.2559B,2012JCAP...09..009W} for the reviews on Fisher forecast and its applications in cosmology).  Using the likelihood $\mathcal{L}$, the Fisher matrix can be defined as:
\begin{eqnarray}
	F_{mn} = -\left \langle \frac{\partial^2\ln\mathcal{L}}{\partial\alpha_m \; \partial\alpha_n}\right \rangle
\end{eqnarray}
where we consider $\{\alpha\}=\left\{\sigma_8, \Omega_m, n_s\right\}$ as the set of model parameters.  ‌Accurate constraining of cosmological parameters using information available at small scales requires modeling the nonlinear effects of matter clustering, galaxy bias, and redshift space distortions (e.g. FoG effect), which are theoretically challenging and here we confine ourselves to use most famous cosmological parameters among the full set of them. A way to overcome these challenges is to use the simulation based inference approach \citep{2016arXiv160506376P,2019MNRAS.488.4440A,2020PNAS..11730055C,2022mla..confE..24H}. Assuming that $\mathcal{L}$ is a multivariate normal distribution, the Fisher matrix element reads as:
\begin{eqnarray}\label{eq:fisher1}
	F_{mn} = \frac{\partial\boldsymbol{\mathcal{D}}^T}{\partial\alpha_m}\; C^{-1} \; \frac{\partial\boldsymbol{\mathcal{D}}}{\partial\alpha_n}
\end{eqnarray}
where $\{\boldsymbol{\mathcal{D}}\}=\{N_{cmd},N_{cr}, \varTheta_{cmd},\varTheta_{cr}\}$ represents the data vector consisting of observables 
and $C$ indicates the covariance matrix. It is worth mentioning that, we use the transformation as $\delta\to \delta^{\prime}=\delta/\sigma_0$, therefore all statistics in the data vector are fully numerically computed
for $\delta^{\prime}$. As an example to compute  the $cmd$‌ statistics, from the density field numerically, we utilize the  discrete form of the volume integrals presented in  Equation (\ref{eq:Ncmd}), which is given by:
\begin{eqnarray}\label{eq:numeric CMD}
	N_{cmd}(\vartheta,i)
	&=&\frac{1}{V} \sum_{p,q,u} \delta_D\left(\delta^{\prime}(p,q,u)-\vartheta\right) \left(\delta^{\prime}_{,i}(p,q,u)\right)^2\; \Delta^3,\nonumber\\
\end{eqnarray}
where $\delta^{\prime}(p,q,u)$ and $\delta^{\prime}_{,i}(p,q,u)$ represent the value and first derivative of the field and in a pixel identified by $(p,q,u)$ indexes in the Cartesian coordinates $(x, y, z)$, respectively. The $\Delta$ represents the pixel size.  We also use the discrete Dirac delta function \citep{schmalzing1997beyond}.‌‌

\begin{figure}
	\includegraphics[width=0.48\columnwidth]{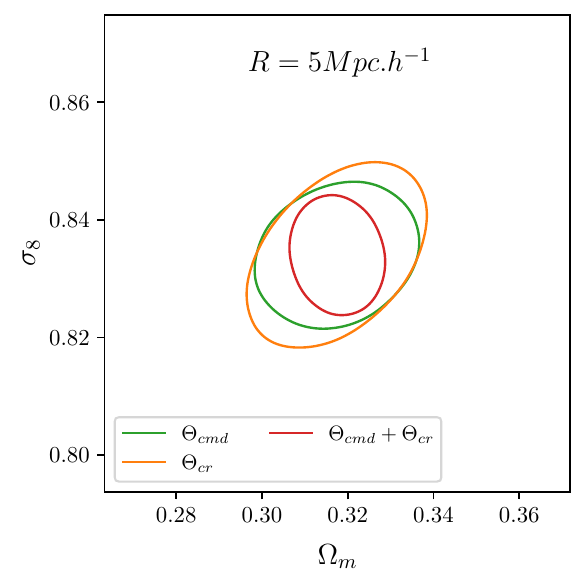}
	\includegraphics[width=0.48\columnwidth]{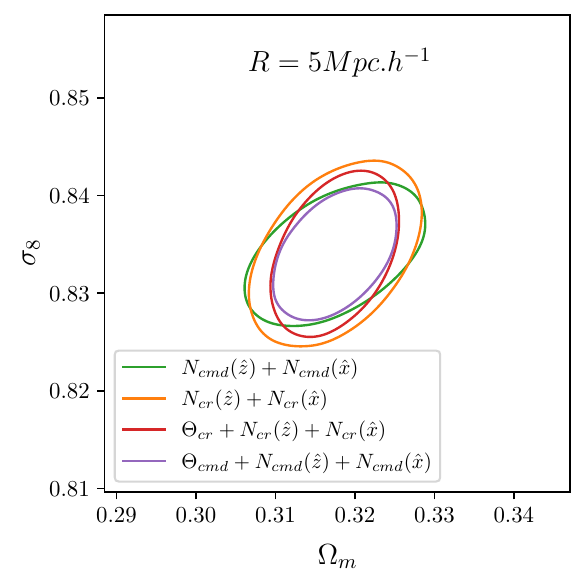}
	\includegraphics[width=0.48\columnwidth]{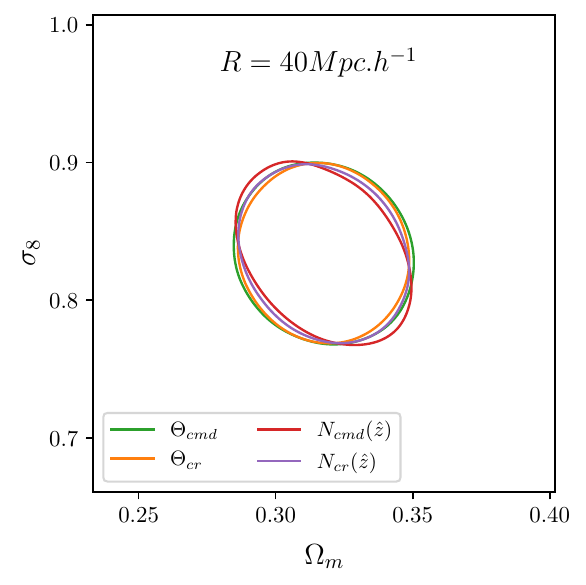}
	\hspace{0.21cm}
	\includegraphics[width=0.48\columnwidth]{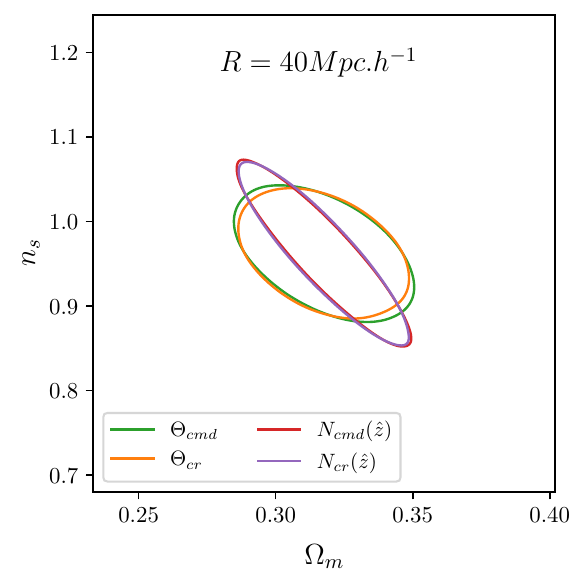}
	\caption{Fisher forecasts: Upper left panel indicates the constraints in the $\Omega_m-\sigma_8$ plane for ratio component, $\varTheta$, and joint analysis of $cmd+cr$. Upper right panel shows the constraints on $\Omega_m$ and $\sigma_8$ by using $N_{cmd}$ and $N_{cr}$ criteria. Lower panels are for $R=40$ Mpc h$^{-1}$, when the non-linear effects are suppressed. In this case the constraint interval on the $\sigma_8$ increases as we expect, while  the impact of $\Omega_m$ due to Kaiser effect remains almost unchanged. Other point is that the $\varTheta_{\diamond}$ can reduce the degeneracy in the $\Omega_m-n_s$ plane respect to $N_{\diamond}$. Al contours have been determined for $68\%$ confidence interval. }\label{fig:fisher1}
\end{figure}

To estimate the partial derivatives in Equation (\ref{eq:fisher1}) for Quijote fiducial values of  cosmological parameters, we consider 500 corresponding realizations. To extract the covariance matrix, we also utilize 5000 realizations of the fiducial simulations. To examine the influence of $cmd$ and $cr$ statistics on parameter constraining, we also takes $R=40$ Mpc h$^{-1}$. Fig. \ref{fig:fisher1} indicates the Fisher forecasts for some relevant parameters. The constraints in the $\Omega_m-\sigma_8$ plane for ratio component ($\varTheta$) and joint analysis of $cmd+cr$ statistics are depicted in the upper left panel. 
Taking into account the $\mathcal{\varTheta}_{cmd}$ instead of $\mathcal{\varTheta}_{cr}$ results in almost $\sim 10\%$ and $\sim20\%$ improvements on constraining the $\Omega_m$ and $\sigma_8$, respectively.  The joint analysis of $\varTheta_{cmd}+\varTheta_{cr}$ also enhances the constraint on the $\sigma_8$ about $35\%$, while for $\Omega_m$ we have $45\%$ compared to $\varTheta_{cr}$ measure. The joint analysis of different components of $cmd$ measures increase the capability to constraint in the $\Omega_m-\sigma_8$ plane.  Incorporating the $R=40$ Mpc h$^{-1}$ and computing the Fisher matrix elements, reveal that constraint interval on the $\sigma_8$ becomes large as we expect, while  the impact of $\Omega_m$ due to Kaiser effect remains almost unchanged. It is worth noting that the $\varTheta_{\diamond}$ can reduce the degeneracy in the $\Omega_m-n_s$ plane respect to the $N_{\diamond}$. A final remark is that, since we have used the unit variance density field, therefore the coefficient of $N_{cmd}$ is independent of $\sigma_8$ similar to  $N_{cr}$, consequently, the constraint on $\sigma_8$ by $N_{cmd}$ for large scale is relatively weak (lower left panel of Fig. \ref{fig:fisher1}).

\begin{figure}
	\centering
	\includegraphics[width=0.8\columnwidth]{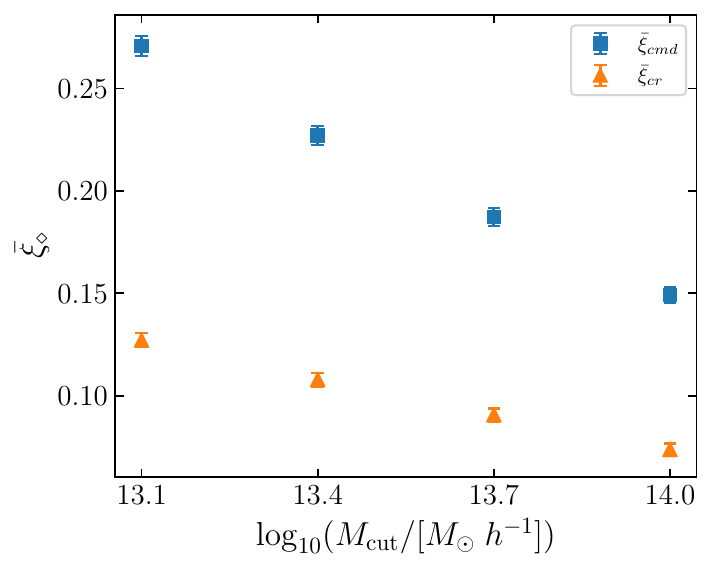}
	\caption{$\bar{\xi}_{(cmd,cr)}$ as a function of $\log_{10}\left(M_{\rm cut}/\left[M_{\odot}{\rm h}^{-1}\right]\right)$ computed from the mock halo catalogs of the Quijote fiducial simulations. We took $R=40$ Mpc h$^{-1}$. }\label{fig:halobias}
\end{figure}

\subsection{Sensitivity to Halo bias}
The visible matters of the Universe forming inside the  gravitationally bound dark matter halo is considered as a representative to trace the dark matter distribution on the cosmological scales. This mechanism inevitably prevents the galaxies perfectly trace the underlying mass distribution. Subsequently, to achieve the proper cosmological inferences by the galaxies observation, the halo bias indicating the relationship between dark matter halos and dark matter distribution on large scale, should be clarified as much as possible \citep{2018PhR...733....1D,2023MNRAS.524.1746L}.\\   
In our approach, the ratio quantity ($\varTheta_{(cmd,cr)}$) depends on $\beta$ according to Equations (\ref{eq:42}) for the Gaussian and linear Kaiser limit. The $\beta$ depends on bias, also, as we defined before, the weighted summation on thresholds $\bar{\xi}_{(cmd,cr)}$ (see Equation (\ref{eq:xibar})) $\bar{\xi}_{(cmd,cr)}$ is related to $\xi_{(cmd,cr)}$ and equivalently to $\varTheta_{(cmd,cr)}$. Subsequently, we expect that by computing the $\bar{\xi}_{(cmd,cr)}$ from the available observational catalogs or from mock data to manage various parameters, in redshift space for different mass cuts is able to reflect the halo bias. To show how our defined parameter in the context of $cmd$ and $cr$ statistics can clarify the halo bias dependency and sensitivity of $cmd$ and $cr$ measures, we calculate $\bar{\xi}_{(cmd,cr)}$ for the mock halo catalogs of the Quijote fiducial simulations to quantify the sensitivity of $N_{cmd}$ and $N_{cr}$ statistics to the halo (or galaxy) bias. we take four mass cuts, $\log_{10}\left(M_{\rm cut}/\left[M_{\odot}{\rm h}^{-1}\right]\right)\in[13.1,13.4,13.7,14.0]$. For each mass cut, we construct a density contrast field  by using halos that have a mass greater than the selected mass cut sampled on a regular lattice with $N_{pix}=256^3$. Then, we smooth the obtained density contrast field using a Gaussian kernel with smoothing scale $R = 40$ Mpc h$^{-1}$. Finally, we calculate the $\bar{\xi}_{cmd}$ and $\bar{\xi}_{cr}$ corresponding to each $M_{\rm cut}$. Fig. \ref{fig:halobias} illustrates the $\bar{\xi}_{\diamond}$ as a function of $\log_{10}\left(M_{\rm cut}/\left[M_{\odot}{\rm h}^{-1}\right]\right)$. The mass cut dependency of $cmd$ and $cr$ according to the Fig.  \ref{fig:halobias} with monotonic behavior means that for a given mass cut value, one can find a unique value for $\bar{\xi}$, consequently the $\bar{\xi}_{\diamond}$ can be considered as a new indicator for examining the  halo (or galaxy) bias dependency. The $1\sigma$ confidence interval for symbols in Fig. \ref{fig:halobias}  decreases by increasing mass cut value. Such behavior can be justified that for the lower value of mass cut, the diversity of halo bias becomes significant leading to obtain the dispersion on averaged halo bias indicated by higher statistical uncertainty. On the contrary, for the higher value of mass cut with statistically enough population, the similarity in the set of derived halo bias increase leading to achieve lower value in the computed statistical error. The results illustrated in Fig. \ref{fig:halobias} have been derived for fiducial simulation and since the $\beta$ depends on the other cosmological parameters in addition to the bias factor, therefore, one cannot yield constraint on the halo (or galaxy) bias. Practically, to mitigate this discrepancy, incorporating the  quantity possessing a lower footprint of the bias should be considered \citep{2020ApJ...896..145A,2021ApJ...907...75A}. In other word, to put constraint on the bias factor which plays as nuisance parameter, we need to know the values of other cosmological parameters such as power spectrum. The $cmd$ is also more sensitive to halo bias compared to $cr$, consequently aiming for constraining on the cosmological parameters in the presence of bias factor considered as a nuisance parameter, the $cr$ statistics is recommended to implement.

\section{Summary and Conclusions}
The redshift space distortions caused by the linear and non-linear effects lead to anisotropy in the density field in the redshift space. To clarify the mention anisotropy as well as non-Gaussianity, we have developed a geometrical measure which is quite sensitive to the anisotropic distribution of density fields.

In this work, inspired by the contour crossing ($cr$) statistic and generalization of MFs, we have introduced the so-called conditional moments of derivative ($cmd$) criteria, which can capture the preferred direction and also are sensitive to the induced anisotropy together with the non-Gaussianity in the underlying cosmological stochastic field. Using a probabilistic framework, we have analytically calculated the theoretical expectation value of $cmd$ measure as a function of threshold ($\vartheta$) for isotropic and anisotropic Gaussian density field in terms of associated spectral indices. Also, for the weakly non-Gaussian field, we have perturbatively extended our analysis up to the $\mathcal{O}(\sigma^2_{0})$ contribution due to the general non-Gaussianity in real and redshift spaces. In addition, to perform a comparison between $cmd$ and $cr$ statistics, we have carried out similar computations for the $cr$ as well. 

Taking into account the Gaussianity and incorporating the linear Kaiser effect as the source of anisotropy in the redshift space density field, we have compared the sensitivity of $cr$ and $cmd$ statistics to the redshift space parameter ($\beta$). The normalized quantity $n_{\diamond}$ depending on direction, threshold, and the $\beta$ parameter has been introduced and our results demonstrated that the $n_{cmd}^{(s)}$ is more sensitive than the $n_{cr}^{(s)}$ to the anisotropy as depicted in the lower panel of Fig.   \ref{fig:example_figure}, particularly for intermediate threshold. According to the error propagation approach and by considering the relative error on the $\varTheta_{\diamond}$ equates to one percent indicating that the $cmd$ enables to put tight constraint compare to other statistics in Gaussian (Fig.   {\ref{Fisher}}) and non-Gaussian regimes.

To make our evaluation more complete, we defined $\xi_{\diamond}$ (Equation   (\ref{eq:xi1})) and its smoothing scale dependency for various values of relevant parameters for the treatment of the influence of FoG and the comparison with the linear Kaiser effect. This quantity is carefully recognized in the range scale where the contribution of FoG or linear Kaiser effect becomes dominant (Fig. \ref{fig:FoG}). This quantity for a high enough value of $R$ also asymptotically goes to the fixed value implying the $\beta$ value.

Implementation of the synthetic data numerically has also supported the good consistency between numerical results and theoretical prediction of $cr$ and $cmd$ measures for the Gaussian field (Fig. \ref{fig:wide_subplot}). To adopt the observational constraining robustly, we also defined a weighted parameter, $\bar{\xi}_{\diamond}$ (Equation (\ref{eq:xibar})), and for various values of $\beta$ in simulation, we revealed that $\bar{\xi}_{cmd}$ has higher $\beta$-dependency (Fig. \ref{fig:xibar}).

Although, we have implemented our methodology on the N-body simulations publicly available by the Quijote suite. The $N^{(r,s)}_{(cmd,cr)}$ as a function of $\vartheta$ implied that there is a deviation from Gaussian theory along the line of sight for both real and redshift spaces and the numerical results are higher(less) than the Gaussian prediction for $\vartheta\gtrsim 0$ ($\vartheta\lesssim 0$) (Fig. \ref{fig:quij11}). To make more sense regarding the non-Gaussianity in the N-body simulations provided by the Quijote, we obtained that the amount of non-Gaussianity in the context of $N_{cmd}$ for perpendicular to the line of sight directions in redshift space is almost the same as for along of line of sight in real space (Fig. \ref{fig:quij12}). This means that to mitigate the non-Gaussianity produced by RSD and to examine non-Gaussianity due to other mechanisms such as primordial ones,  we should consider the analysis on the plane perpendicular to the line of sight in redshift space.  The peculiar velocity magnifies the non-Gaussianity along the line of sight in redshift space which is well recognized by $cmd$ and $cr$ measures.

To quantify the constraining power of $cmd$ and $cr$ measured, we have done Fisher forecasts. Numerically ‌determining the associated matrix elements clarified the influences of our statistical measures individually accompanying the joint analysis on the relevant cosmological parameters (Fig. (\ref{fig:fisher1})). We achieved that constraints on $\sigma_8$ and $\Omega_m$ according to joint analysis of $\varTheta_{cmd}+\varTheta_{cr}$ are better by $35\%$ and $45\%$ relative to considering $\varTheta_{cr}$, respectively. Taking into account the ratio quantity of our measures leads to reduce the degeneracy in the $\Omega_m-n_s$ plane compared to that of given by $N_{(cmd,cr)}$. 

We have also attempted to address the sensitivity of the $cmd$ and $cr$ measures to the halo bias in redshift space. Using the Quijote halo catalogs, we have calculated the $\bar{\xi}_{(cmd,cr)}$ for the density field constructed from halos whose masses are greater than a specific mass cut (Fig. \ref{fig:halobias}). Our results confirmed that $\bar{\xi}$ can be a promising measure to evaluate halo bias. While the $cr$ measure with lower dependency on mass cut revealed robust quantity when the bias factor is considered as a nuisance parameter.

To go further, we suggest to do following tasks as the complementary subjects in the banner of excursion sets and RDS and will be left for the future study: however the plane-parallel approximation provides a reliable approach in accounting for the peculiar velocity along the line of sight in the observed distribution of galaxies, but to obtain more accurate evaluation, spherical redshift distortions could be interesting to pursue. We also focused on the matter density field and it is useful to consider the galaxies catalogs and other real data sets instead, consequently utilizing the $cmd$ and $cr$ measures open new room to evaluate bias factor. Utilizing the $cmd$ and $cr$ for cosmological parameters constraining approach need to do more complementary analysis like assessing the  parameters associated with nonlinear phenomena such as $\sigma_u$, and a recommended method is to use the simulation based inference approach \citep{2016arXiv160506376P,2019MNRAS.488.4440A,2020PNAS..11730055C,2022mla..confE..24H}. In addition, various model of primordial non-Gaussianity can be evaluated.

Generally, the existence of a preferred direction in cosmology and for various scales has remained under debate topic, to this end, our measures can provide a pristine framework. Thanks to scaling window analysis and modifying the $cmd$ criteria, hopefully, makes it more capable by scanning over the underlying field to capture the scale and location dependency of directional behavior   \citep{li2013detection,nezhadhaghighi2015crossing,klatt2022characterization,2010JSMTE..11..010K,2013NJPh...15h3028S}.

\acknowledgments

   The authors are very grateful to Ali Haghighatgoo for his extremely useful comments on different parts of this paper. Also thanks to Ravi K. Sheth for his constructive discussions. SMSM appreciates the hospitality of the HECAP section of ICTP where a part of this research was completed. We also thank the Quijote team for sharing its simulated data sets and providing extensive instruction on how to utilize the data. Finally, we appreciate the anonymous referee who helped us to focus on the most relevant topics which led to improved our paper.

 \begin{appendices}
\section{Generalization of Minkowski Functionals}	
 	 The so-called Minkowski Functionals (MFs) possess scalar property. To characterize the morphology of a typical $d$-dimensional field, there are $(d+1)$ MFs which are unique and complete in the sense of {\it Hadwigers's theorem} and satisfying the motion invariance (e.g. rotations and translations), additivity and conditional continuity. It is well-known that the additivity and motion-invariance properties of the MFs lead to prevent the MFs from discriminating different anisotropic patterns in a field \citep{Beisbart2002}. Depending on starting point objective which is in principle devoted to the mathematical side as well as from the applications, substantial progress can be considered to generalize scalar MFs. Relaxing the above conditions allows to consider the following generalization of the MFs in $d-$dimension: 
 	\begin{eqnarray}\label{eq:MVs}
 	\Xi &\equiv&\frac{1}{V_d}\int_{V_d}dV_d\;\mathcal{G}(s_{\nu};{\vec{r}},\delta,\boldsymbol{\nabla}\delta,...)
 	\end{eqnarray}
 	where $s_{\nu}$ is a functional form of curvatures and $\nu=0,...,(d-1)$. The $\mathcal{G}$ is a general functional form of $(s_{\nu};{\boldsymbol{r}},\delta,\boldsymbol{\nabla}\delta,...)$. A reasonable extension of scalar MFs on Euclidean space has been done by introducing a specific functional form for $\mathcal{G}$ which is known as the ``Minkowski valuations'' (MVs)
 	\citep{McMullen1997,Alesker1999,Hug2007TheSO}. In this regard, we have: 
 	\begin{eqnarray}\label{eq:MVscom}
 	\begin{split}
 	\mathcal{W}_{\nu}^{(p,q)} \equiv&\frac{1}{V_d}\int_{\partial Q_{\vartheta}}\;dA_d\; s_{\nu} \overbrace{{\boldsymbol{r}}\otimes{\boldsymbol{r}}\otimes...\otimes {\boldsymbol{r}}}^{p-times}\\
 	&\otimes\; \underbrace{\frac{\boldsymbol{\nabla}\delta}{|\boldsymbol{\nabla}\delta|}\otimes\frac{\boldsymbol{\nabla}\delta}{|\boldsymbol{\nabla}\delta|}\otimes ...\otimes \frac{\boldsymbol{\nabla}\delta}{|\boldsymbol{\nabla}\delta|}}_{q-times}
 	\end{split}
 	\end{eqnarray}
 	here $\otimes$ reveals the tensor product. Accordingly, the vectorial form is derived for $(p=1,q=0)$, while the $\mathcal{W}_{\nu}^{(p,q)}$ for $(p=0,q=1)$ by definition is vanished. Also for rank-2 tensor form, the condition $p+q=2$ should be satisfied in Equation (\ref{eq:MVscom}).

\end{appendices}	 
   

\end{document}